\documentclass[journal,final,twocolumn,10pt,twoside]{IEEEtran}

\normalsize

\usepackage{graphicx,epsfig,epstopdf,amsmath,amssymb,enumerate,color}
\usepackage{cite}  
\usepackage{algpseudocode}

\ifCLASSINFOpdf
\else
\fi

\begin{document}
\title{A General 3D Non-Stationary 5G Wireless Channel Model}

\author{Shangbin~Wu, Cheng-Xiang~Wang,~\IEEEmembership{Fellow,~IEEE,}  el-Hadi~M.~Aggoune,~\IEEEmembership{Senior Member,~IEEE,} Mohammed M. Alwakeel,~\IEEEmembership{Senior Member,~IEEE,} and Xiao-Hu~You,~\IEEEmembership{Fellow,~IEEE}

\thanks{Manuscript received August 29, 2017; revised November 12, 2017; accepted November 24, 2017. The authors gratefully acknowledge the support of this work from the EU H2020 ITN 5G Wireless project (No. 641985), the EU H2020 RISE TESTBED project (No. 734325), the EU FP7 QUICK project (No. PIRSES-GA-2013-612652), the EPSRC TOUCAN project (No. EP/L020009/1), the Natural Science Foundation of China (No. 61210002), and SNCS Research Center, the University of Tabuk, Saudi Arabia. Part of this work was presented in Dr. S. Wu's PhD thesis, October 2015. The associate editor coordinating the review of this paper and approving it for publication was L. Dai.}% <-this % stops a space

\thanks{S. Wu is with Samsung R\&D Institute UK, Staines-upon-Thames, TW18 4QE, UK (E-mail: shangbin.wu@samsung.com).}

\thanks{C. -X. Wang (corresponding author) is with the Institute of Sensors, Signals and Systems, School of Engineering and Physical Sciences, Heriot-Watt University, Edinburgh, EH14 4AS, U.K. He is also with the Mobile Communications Research Laboratory, Southeast
University, Nanjing, 211189, China (E-mail: cheng-xiang.wang@hw.ac.uk).}% <-this % stops a space
\thanks{el-H. M. Aggoune and M. M. Alwakeel are with the Sensor Networks and Cellular Systems (SNCS) Research Center, University of Tabuk, P. O. Box: 6592-2, 47315/4031 Tabuk, Saudi Arabia (E-mail: \{haggoune.sncs, alwakeel\}@ut.edu.sa).}
\thanks{X. -H. You is with the Mobile Communications Research Laboratory, Southeast
University, Nanjing, 211189, China (E-mail: xhyu@seu.edu.cn).}% <-this % stops a space

%\thanks{}
%\thanks{TCOM version based on Michael Shell's bare{\textunderscore}jrnl.tex version 1.3.}
}


\markboth{IEEE TRANSACTIONS ON COMMUNICATIONS, vol. xx, no. xx, month 2017}%
{Submitted paper}



\maketitle
%\vspace{-2.2cm}
\begin{abstract}

%\boldmath
A novel unified framework of geometry-based stochastic models (GBSMs) for the fifth generation (5G) wireless communication systems is proposed in this paper. The proposed general 5G channel model aims at capturing small-scale fading channel characteristics of key 5G communication scenarios, such as massive multiple-input multiple-output (MIMO), high-speed train (HST), vehicle-to-vehicle (V2V), and millimeter wave (mmWave) communication scenarios. It is a three-dimensional (3D) non-stationary channel model based on the WINNER II and Saleh-Valenzuela (SV) channel models considering array-time cluster evolution. Moreover, it can easily be reduced to various simplified channel models by properly adjusting model parameters. Statistical properties of the proposed general 5G small-scale fading channel model are investigated to demonstrate its capability of capturing channel characteristics of various scenarios, with excellent fitting to some corresponding channel measurements.
\end{abstract}


% \fi
\IEEEpeerreviewmaketitle

\section{Introduction}
\IEEEPARstart{T}{o} satisfy the demands of the fifth generation (5G) wireless communication networks, known as increased data rate, reduced latency, energy, and cost \cite{Andrews14}, a number of advanced technologies have been proposed in the literature as potential 5G technologies.
Massive multiple-input multiple-output (MIMO), i.e., an enhanced MIMO technique with a large number of antennas, is able to greatly improve communication reliability, spectral efficiency, and energy efficiency \cite{MassiveMIMO4NGWS}--\hspace{-0.001cm}\cite{Lu14}. Vehicle-to-vehicle (V2V) communications \cite{Wang09}, \cite{Wang14} were proposed in the heterogeneous network architecture of 5G to enable vehicles to connect mutually without base stations.
High-speed train (HST) communication also attracts attention for the emerging development of high mobility trains with speed expected to be higher than $500$ km/h.
Furthermore, millimeter wave (mmWave) frequency bands (30--300 GHz) have been proposed to be used for 5G wireless communications to solve the spectrum crisis problem.
The mmWave frequency bands are capable of providing large bandwidth (in the order of GHz) and exploiting polarization and massive MIMO \cite{Pi11}--\hspace{-0.001cm}\cite{Brady13}.
In order to design and evaluate 5G systems, a channel model which can capture channel characteristics of the above-mentioned potential technologies is essential.
However, conventional channel models such as 3GPP spatial channel model (SCM) \cite{25996}, the WINNER II \cite{winner}, WINNER+ \cite{winner+}, 3GPP three dimensional (3D) SCM \cite{36873}, IMT-A \cite{IMTA}, and COST 2100 channel models \cite{WC_COST2100}--\hspace{-0.001cm}\cite{Pervasive} are not able to sufficiently meet these emerging 5G requirements.
Although both large-scale fading and small-scale fading have large impacts on system performance, small-scale fading of 5G wireless channels is less studied in the literature.

The first geometry-based stochastic model (GBSM) for 5G wireless channels was proposed by the METIS project \cite{METIS}.
However, the METIS GBSM did not sufficiently support channel characteristics of massive MIMO, V2V, and mmWave communications \cite[Table 4--1]{METIS}.
Later, multiple industrial partners and academic institutes formed a special interest group (SIG) and jointly published a white paper for 5G channel models \cite{5GCM}. This SIG white paper has covered new channel characteristics from 6 GHz to 100 GHz, such as blockage, spatial consistency, support of large bandwidth and array, and novel path loss models.
The SIG white paper was then used as the guideline for the 3GPP new radio (NR) channel model for frequencies from 0.5 to 100 GHz in \cite{38901}.
Based on the IMT-A channel model and 3GPP 3D SCM, the IMT-2020 channel model \cite{IMT-2020} was recently proposed by the ITU.
The IMT-2020 channel model supports frequency bands up to 100 GHz and can cover many new features, e.g., 3D propagation, spatial consistency, large bandwidth, large antenna array, etc.
Additionally, another modeling approach known as the map-based model was introduced in the METIS channel model \cite{METIS}, 3GPP NR channel model \cite{38901}, and IMT-2020 channel model \cite{IMT-2020}.
Map-based models aim at computing channel coefficients in a deterministic manner when the layout of a network is predefined. The millimetre-wave evolution for backhaul and access (MiWEBA) project \cite{MiWEBA} proposed a channel model combining deterministic approach based on network layout with stochastic components to model mmWave channels.
However, since layouts of networks are not always accessible, GBSMs attract more attentions from researchers. Therefore, we will focus on GBSMs and small-scale fading in this paper.
%
%Although the METIS project has also proposed a map-based model to address these features, map-based models are less used in standard bodies such as 3GPP. Since GBSMs have widely been used to model small-scale channel fading in the WINNER II \cite{winner}, 3GPP spatial channel model \cite{36211}, IMT-A \cite{IMTA}, and COST 2100 channel models \cite{WC_COST2100}--\hspace{-0.001cm}\cite{Pervasive},

\subsection{Related Work I: GBSMs for Massive MIMO}
It was reported in \cite{2dot6GHz} and \cite{Gao12} that massive MIMO channels have some specific characteristics which were ignored in conventional MIMO (numbers of transmit and receive antennas are relatively small) channel models \cite{winner}--\hspace{-0.001cm}\cite{Pervasive}.
First, when the number of antennas is large, the distance $\mathcal{D}$ between the transmitter (cluster) and the receiver may not be larger than the Rayleigh distance $2\mathcal{L}^2/\lambda$ \cite{Saunders07}, where $\mathcal{L}$ is the dimension of the antenna array and $\lambda$ is the carrier wavelength. In this case, the effect of spherical wavefront is significant and the conventional plane wavefront assumption in \cite{winner}--\hspace{-0.001cm}\cite{IMTA} is not well-justified.
The COST 2100 channel model and the METIS GBSM can support spherical wavefronts after minor extensions.
Second, cluster appearance and disappearance can occur on the array axis. As a result, each antenna may have its own set of observable clusters.
This has not been considered in most advanced GBSMs such as the SCM-extension model \cite{Baum05}, the COST 2100 channel model \cite{WC_COST2100}, the METIS GBSM \cite{METIS}, the 3GPP NR channel model \cite{38901}, and the IMT-2020 channel model \cite{IMT-2020}.
A two-dimensional (2D) confocal ellipse model and a 3D twin-cluster model were proposed for massive MIMO channels in \cite{Wu15} and \cite{Wu14}, respectively, while a comprehensive survey of massive MIMO channel measurements and models was given in \cite{Wang16SCIS}.
Both the ellipse model and twin-cluster model in \cite{Wu15} and \cite{Wu14} employed the spherical wavefront assumption and adapted the cluster birth-death process in \cite{Zwick00} and \cite{Zwick02} to both the time and array axes to characterize cluster appearance and disappearance.
However, the mean power evolution of clusters and rays, directional antennas, and polarized antennas were ignored in \cite{Wu15} and \cite{Wu14}.
Most importantly, channel models in \cite{Wu15} and \cite{Wu14} were not designed to accommodate mmWave channels, V2V channels, and arbitrary antenna array layouts.

\subsection{Related Work II: GBSMs for V2V and HST}
In V2V channels, the transmitter, scatterers, and receiver can be all moving. Doppler frequencies caused by either the transmitter and/or scatterers and/or receiver should be taken into account. Wideband 2D GBSMs for V2V channels were proposed and validated via measurements in \cite{Cheng13}--\hspace{-0.001cm}\cite{Karedal09}, where clusters were categorized into mobile clusters and static clusters. In V2V communications, transmitters and receivers may be lower than clusters on surrounding buildings. Therefore, 3D clusters were included in \cite{Zajic09}--\hspace{-0.01cm}\cite{Yuan14}. A 3D concentric-cylinder V2V channel model was introduced in \cite{Zajic09}, while 3D GBSMs combining a two-sphere model and an elliptic-cylinder model were proposed in \cite{Yuan15} and \cite{Yuan14}.
Recently, birth-death process was used to model cluster dynamics in V2V channels in \cite{He15}. For HST communications, relevant non-stationary GBSMs can be found in \cite{Ghazal15}--\hspace{-0.001cm}\cite{Liu17} and some measurement results were given in \cite{Chen12}. The non-stationary channel behavior of HST systems is similar to that of V2V channels with large Doppler frequencies. However, cluster evolution on the time axis was ignored in the METIS GBSM \cite{METIS}, 3GPP NR channel model \cite{38901}, IMT-2020 channel model \cite{IMT-2020}, and models in \cite{Cheng13}--\hspace{-0.001cm}\cite{He15}. Thus, it is difficult to track the channel with respect to time in a continuous manner.

\subsection{Related Work III: GBSMs for mmWave}
As the supported bandwidth for mmWave is large (in the order of GHz \cite{Pi11}--\hspace{-0.001cm}\cite{Brady13}), GBSMs for mmWave channels need to consider high delay resolution, i.e., rays within a cluster may be resolvable and the numbers of rays within clusters may vary.
The GBSM of the METIS channel model \cite{METIS} supports frequency bands up to 70 GHz. However, the resolvable rays within clusters and the varying numbers of rays within clusters were not included.
The quasi deterministic radio channel generator (QuaDRiGa) channel model \cite{Jaeckel14} was used as the initial model for mmWave channels in the mmWave based mobile radio access network for 5G integrated communications (mmMAGIC) project \cite{mmMAGIC}.
However, it also ignored the resolvable rays and varying numbers of rays within clusters.
The Saleh-Valenzuela (SV) channel model \cite{Saleh87}, which was originally proposed for indoor multipath propagations, has been used to evaluate system performance in the IEEE wireless personal area network (PAN) standard \cite{802154a}--\hspace{-0.001cm}\cite{Molisch03} with the supported bandwidth over $500$ MHz.
Since mmWave channels are expected to have bandwidths over $500$ MHz, applications of the SV channel model to mmWave channels can be found in \cite{Maltsev10}--\hspace{-0.001cm}\cite{Huang17}.
In a SV channel model, the number of rays within each cluster was assumed to follow a Poisson distribution. Complex gain and delay were assigned to each ray. A 3D mmWave channel model with random numbers of clusters and paths within each cluster was proposed in \cite{Samimi15}. However, in the above SV-based mmWave channel models, time evolution and mean ray power evolution have not sufficiently been studied.

\subsection{Contributions}
To address the abovementioned research gaps, this paper has the following contributions \cite{WuS15PhD}.
\begin{enumerate}
\item A general 3D non-stationary 5G GBSM for terrestrial wireless communication systems is proposed, having the capability of simulating massive MIMO, V2V, HST, and mmWave small-scale fading channels \cite{WuS15PhD}. It also considers time evolution of channels with all the model parameters as time varying. The proposed general 5G channel model is based on the WINNER II channel model \cite{winner}, in order to keep consistency of 4G and 5G channel models, and the SV channel model in order to support the high delay resolution of mmWave channels \cite{802154a}--\hspace{-0.001cm}\cite{Molisch03}. Spherical wavefront and array-time cluster evolution are included to represent massive MIMO channel characteristics. Array-time evolution includes cluster birth-death process in both the array and time axes and geometrical relationship updates. The mean power updates of rays are also embedded in the proposal 5G small-scale fading channel model with the assumption of the inverse square law.
%
%\item Array-time evolution is considered in the proposed unified 5G GBSM framework. This includes cluster birth-death process in both the array and time axes and geometrical relationship updates. The mean power updates of rays are also embedded in the proposal 5G channel model with the assumption of inverse square law.

\item The proposed general 5G GBSM can easily be reduced to various simplified channel models by setting proper channel parameters, which is demonstrated by fitting some statistical properties to the corresponding channel measurement data.

\end{enumerate}

The rest of this paper is organized as follows. Section~\ref{sec_Unified_Framework} gives a general description of the proposed unified framework for 5G small-scale fading channel models. Statistical properties of the proposed general 3D 5G GBSM are investigated in Section~\ref{sec_StochasticPropertyAnalysis}. Simulation/numerical results and analysis are presented in Section \ref{sec_numerical_analysis_section}. Conclusions are drawn in Section~\ref{sec_conclusion_section}.

%%%%%%%%%%%Introduction ends%%%%%%%%%%%%%%

%%%%%%%%%%%{Unified Framework for 5G Channel Models begins%%%%%%%%%%%%
\section{A General 3D Non-Stationary 5G Small-Scale Fading Channel Model} \label{sec_Unified_Framework}
\begin{figure}
\centering\includegraphics[width=3.5in]{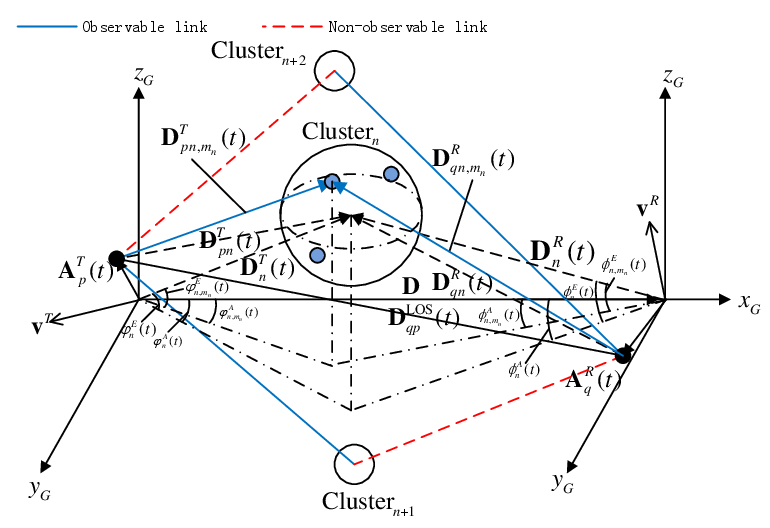}
\caption{A general 3D non-stationary 5G GBSM.}
\label{fig_Unified_framework}
\end{figure}

Let us consider a MIMO system with $M_R$ receive and $M_T$ transmit antennas communicating at carrier frequency $f_c$. Let $\mathrm{Ant}_q^R$ denote the $q$th receive antenna and $\mathrm{Ant}_p^T$ denote the $p$th transmit antenna. Also, let $\mathrm{Cluster}_n$ denote the $n$th cluster. It should be noted that arbitrary antenna array layouts are assumed in the proposed model. Typical antenna array layouts include uniform linear arrays, 2D planar arrays, and 3D cube arrays. Antenna responses can be modified subject to actual antenna settings. The scattering environment between the transmitter and receiver is abstracted as effective clusters \cite{winner}, which characterize the first and last bounces of the channel. Multi-bounces between the first and last bounces are abstracted by a virtual link. The proposed general 3D non-stationary 5G GBSM is illustrated in Fig. \ref{fig_Unified_framework}. It should be noticed that ($x_{\mathrm{G}},y_{\mathrm{G}},z_{\mathrm{G}}$) axes are established as the global coordinate system (GCS) with origin at the center of the transmit array. This needs to be distinguished from the local coordinate systems (LCSs) with origins at the centers of transmit and receive arrays when calculating antenna pattern in a 3--D space. Spherical wavefront and cluster appearance and disappearance are assumed in order to support massive MIMO scenarios. In this case, each antenna may have its own set of observable clusters. Let $C^R_q(t)$ and $C^T_p(t)$ represent the cluster sets of $\mathrm{Ant}_q^R$ and $\mathrm{Ant}_p^T$ at time $t$, respectively. Then, the total number of clusters $N(t)$ observable by both the transmitter and receiver at time $t$ can be calculated as
$N(t)= \textrm{card}\left(\bigcup\limits ^{M_T}_{p=1} \bigcup\limits^{M_R}_{q=1}S_{qp}(t)\right)$
%\begin{equation}
%N(t)= \textrm{card}\left(\bigcup\limits ^{M_T}_{p=1} \bigcup\limits^{M_R}_{q=1}S_{qp}(t)\right)
%\label{equ_N}
%\end{equation}
where $S_{qp}(t)=C^R_q(t)\bigcap C^T_p(t)$, $\mathrm{card}(\mathcal{S} )$ denotes the cardinality of the set $\mathcal{S}$, $\bigcup$ and $\bigcap$ denote the union and intersection of sets, respectively.
Also, both the transmitter and receiver are assumed to be in motion to support V2V scenarios, which results in Doppler frequencies at both sides. In addition, rays within clusters are considered to be resolvable in order to support the high time resolution in mmWave scenarios. Complex gain and delay should be assigned to each ray.

%\begin{figure*}[!t]
%% ensure that we have normalsize text
%\normalsize
%% Store the current equation number.
%% Set the equation number to one less than the one
%% desired for the first equation here.
%% The value here will have to changed if equations
%% are added or removed prior to the place these
%% equations are referenced in the main text.
%\begin{align}
%R_{1}(\rho)&=\int\limits_0^{\infty}\log_2(1+\rho\xi)dF_{\Xi}(\xi)=\int\limits_0^{\infty}\log_2(1+\rho\xi)\frac{1}{(M_R-1)!}\xi^{M_R-1}e^{-\xi}d\xi\nonumber\\
%&=\left\lbrace \begin{matrix}
%-\frac{1}{\ln 2}\mathrm{Ei}\left(-\frac{1}{\rho} \right)\exp\left(\frac{1}{\rho} \right) & (M_R=1)
%\\
%\frac{1}{\ln 2}\sum\limits_{m=0}^{M_R-1}\frac{1}{(M_R-1-m)!}\left[\frac{(-1)^{M_R-m}}{\rho^{M_R-1-m}}\mathrm{Ei}\left(-\frac{1}{\rho} \right)\exp\left(\frac{1}{\rho} \right)+\sum\limits_{k=1}^{M_R-1-m}\frac{(k-1)!}{(-\rho)^{M_R-1-m-k}}  \right] & (M_R> 1),
%\end{matrix}\right.
%\label{equ_instant_sumrate_low}
%\end{align}
%% Restore the current equation number.
%% IEEE uses as a separator
%\hrulefill
%% The spacer can be tweaked to stop underfull vboxes.
%\vspace*{4pt}
%\end{figure*}

\begin{table*}[ht]
\caption{Definitions of key 5G channel model parameters.}
\center
\footnotesize

    \begin{tabular}{|c|c|}
    \hline
    $\psi_A^{R}$, $\psi_E^{R}$  & azimuth and elevation angles of the receive array broadside  \\ \hline
    $\psi_A^{T}$, $\psi_E^{T}$  & azimuth and elevation angles of the transmit array broadside  \\ \hline
        $N(t)$  & total number of observable clusters  \\ \hline
    $M_n$  & number of rays within $\mathrm{Cluster}_n$  \\ \hline
    $C^R_q(t)$, $C^T_p(t)$  & cluster set of $\mathrm{Ant}_q^{R}$ and cluster set of $\mathrm{Ant}_p^{T}$ \\ \hline
        $S_{qp}(t)$  & set of clusters observable for both $\mathrm{Ant}_q^{R}$ and $\mathrm{Ant}_p^{T}$ \\ \hline

    $\phi_n^{A}(t)$, $\phi_n^{E}(t)$  &  azimuth and elevation angles between $\mathrm{Cluster}_n$ and the receive array center \\ \hline
    $\varphi_n^{A}(t)$, $\varphi_n^{E}(t)$  &  azimuth and elevation angles between $\mathrm{Cluster}_n$ and the transmit array center \\ \hline
    $\phi_{n,m_n}^{A}(t)$, $\phi_{n,m_n}^{E}(t)$  &  azimuth and elevation angles between the $m_n$th ray of $\mathrm{Cluster}_n$ and the receive array center  \\ \hline
    $\varphi_{n,m_n}^{A}(t)$, $\varphi_{n,m_n}^{E}(t)$  &  azimuth and elevation angles between the $m_n$th ray of $\mathrm{Cluster}_n$ and the transmit array center \\ \hline
    $\mathbf{A}_q^R (t)$, $\mathbf{A}_p^T (t)$ & 3D position vectors of $\mathrm{Ant}_q^{R}$ and $\mathrm{Ant}_p^{T}$  \\ \hline
    $\mathbf{D}_n^R (t)$, $\mathbf{D}_n^T (t)$ &3D distance vectors between $\mathrm{Cluster}_n$ and the receiver (transmitter) array center  \\ \hline
    $\mathbf{D}_{n,m_n}^R (t)$,$\mathbf{D}_{n,m_n}^T (t)$ &3D distance vectors between $\mathrm{Cluster}_n$ and the receive (transmit) array center via the $m_n$th ray  \\ \hline
    $\mathbf{D}_{qn,m_n}^R (t)$,$\mathbf{D}_{pn,m_n}^T (t)$ &3D distance vectors between $\mathrm{Cluster}_n$ and $\mathrm{Ant}_q^{R}$ ($\mathrm{Ant}_p^{T}$) via the $m_n$th ray  \\ \hline
    $f_{qn,m_n}^R (t)$,$f_{pn,m_n}^T (t)$ & Doppler frequencies of $\mathrm{Ant}_q^{R}$ ($\mathrm{Ant}_p^{T}$) via $\mathrm{Cluster}_n$ and the $m_n$th ray  \\ \hline
    $\mathbf{D}_{qp}^{\mathrm{LOS}} (t)$ &3D distance vector of the LOS component between $\mathrm{Ant}_q^{R}$ and $\mathrm{Ant}_p^{T}$  \\ \hline
    $f_{pq}^\mathrm{LOS} (t)$& Doppler frequency of the LOS component between $\mathrm{Ant}_q^{R}$ and $\mathrm{Ant}_p^{T}$  \\ \hline
    $\mathbf{v}^R$,$\mathbf{v}^T$ &3D velocity vectors of the receive and transmit arrays  \\ \hline
    $\mathbf{v}_n^R$, $\mathbf{v}_n^T$ &3D velocity vectors of the last bounce and first bounce of $\mathrm{Cluster}_n$  \\ \hline
    $P_{n,m_n}(t)$ & mean power of the $m_n$th ray of $\mathrm{Cluster}_n$  \\ \hline
    $\mathbf{D}$ &3D distance vector between the receive and transmit array centers  \\ \hline
$K$ & Rician factor  \\ \hline

$\tilde{\lambda} $ & mean number of rays within a cluster  \\ \hline
${\kappa} $ &  cross polarization power ratio   \\ \hline
${\lambda_G,\lambda_R} $ & generation rate and recombination rate of clusters  \\ \hline

    \end{tabular}
    \label{tab_key_geometry}
\end{table*}
%Here, we define the central azimuth angle as the angle between the projection of the line segment connecting an object and array center on the $x$--$y$ plane and the positive $x$-axis, and define the central elevation angle as the complement angle between the projection of the line segment connecting an object and array center on the $x$--$y$ plane and the positive $z$-axis. Given the azimuth and elevation angles at both the receiver and transmitter sides,

Let $\mathbf{A}^R_q(t)$ and $\mathbf{A}^T_p(t)$ denote the position vectors of $\mathrm{Ant}_q^{R}$ and $\mathrm{Ant}_p^{T}$, respectively. Also, let $\psi_A^{R}$ and $\psi_E^{R}$ be azimuth and elevation angles of the receive array broadside, and let $\psi_A^{T}$ and $\psi_E^{T}$ be azimuth and elevation angles of the transmit array broadside, respectively.
%\begin{equation}
%\mathbf{A}^R_q(t)=\frac{M_R-2q+1}{2}\delta_R\begin{bmatrix}
%\cos \upsilon^R_E(t) \cos\upsilon^R_A(t)\\
%\cos \upsilon^R_E(t) \sin\upsilon^R_A(t)\\
%\sin\upsilon^R_E(t)
%\end{bmatrix}^{\mathrm{T}}+\mathbf{D}
%\label{equ_A_R_q}
%\end{equation}
%\begin{equation}
%\mathbf{A}^T_p(t)=\frac{M_T-2p+1}{2}\delta_T\begin{bmatrix}
%\cos \upsilon^T_E(t) \cos\upsilon^T_A(t)\\
%\cos \upsilon^T_E(t) \sin\upsilon^T_A(t)\\
%\sin\upsilon^T_E(t)
%\end{bmatrix}^{\mathrm{T}}
%\label{equ_A_T_p}
%\end{equation}
Let $\mathbf{D}$ denote the initial position vector of the receiver and is assumed to equal $\left[ D,0,0\right]^{\mathrm{T}}$, and $D$ is the initial distance between the transmitter and receiver centers. The line-of-sight (LOS) distance vector $\mathbf{D}^{\mathrm{LOS}}_{qp}(t)$ between $\mathrm{Ant}_q^{R}$ and $\mathrm{Ant}_p^{T}$ is computed as
\begin{equation}
\mathbf{D}^{\mathrm{LOS}}_{qp}(t)=\mathbf{A}^R_{q}(t)-\mathbf{A}^T_{p}(t).
\label{equ_D_qp_LOS}
\end{equation}
According to the geometrical relationships in Fig. \ref{fig_Unified_framework} and the key parameters listed in Table \ref{tab_key_geometry}, distance vectors of $\mathrm{Cluster}_n$ at the transmitter and receiver are calculated as
\begin{equation}
\mathbf{D}^R_{n}(t)=D^R_n(t)\begin{bmatrix}
\cos \phi^E_{n}(t) \cos\phi^A_{n}(t)\\
\cos \phi^E_{n}(t) \sin\phi^A_{n}(t)\\
\sin\phi^E_{n}(t)
\end{bmatrix}^{\mathrm{T}}+\mathbf{D}
\label{equ_D_R_n}
\end{equation}
\begin{equation}
\mathbf{D}^T_{n}(t)=D^T_n(t)\begin{bmatrix}
\cos \varphi^E_{n}(t) \cos\varphi^A_{n}(t)\\
\cos \varphi^E_{n}(t) \sin\varphi^A_{n}(t)\\
\sin\varphi^E_{n}(t)
\end{bmatrix}^{\mathrm{T}}
\label{equ_D_T_n}
\end{equation}
where $D^R_n(t)$ and $D^T_n(t)$ are the Frobenius norms of $\mathbf{D}^R_{n}(t)$ and $\mathbf{D}^T_{n}(t)$, respectively. Distance vectors of the $m_n$th ray of $\mathrm{Cluster}_n$ to the transmitter and receiver centers are calculated as
\begin{equation}
\mathbf{D}^R_{n,m_n}(t)=D^R_n(t)\begin{bmatrix}
\cos \phi^E_{n,m_n}(t) \cos\phi^A_{n,m_n}(t)\\
\cos \phi^E_{n,m_n}(t) \sin\phi^A_{n,m_n}(t)\\
\sin\phi^E_{n,m_n}(t)
\end{bmatrix}^{\mathrm{T}}+\mathbf{D}
\label{equ_D_R_n_m_n}
\end{equation}
\begin{equation}
\mathbf{D}^T_{n,m_n}(t)=D^T_n(t)\begin{bmatrix}
\cos \varphi^E_{n,m_n}(t) \cos\varphi^A_{n,m_n}(t)\\
\cos \varphi^E_{n,m_n}(t) \sin\varphi^A_{n,m_n}(t)\\
\sin\varphi^E_{n,m_n}(t)
\end{bmatrix}^{\mathrm{T}}.
\label{equ_D_T_n_m_n}
\end{equation}
Distance vectors between the $m_n$th ray of $\mathrm{Cluster}_n$ and antenna elements are calculated as
\begin{equation}
\mathbf{D}^R_{qn,m_n}(t)=\mathbf{D}^R_{n,m_n}(t)-\mathbf{A}^R_{q}(t)
\label{equ_D_R_qn_m_n}
\end{equation}
\begin{equation}
\mathbf{D}^T_{pn,m_n}(t)=\mathbf{D}^T_{n,m_n}(t)-\mathbf{A}^T_{p}(t).
\label{equ_D_T_pn_m_n}
\end{equation}
It should be noticed that position vectors are all time dependent. After all vectors are obtained in the 3D space, the channel impulse response can be derived.
\subsection{Channel Impulse Response}
Based on the WINNER II and SV channel models, the proposed 5G GBSM at time $t$ with delay $\tau$ can be characterized by an $M_R\times M_T$ matrix $\mathbf{H}(t,\tau)=\left[h_{qp}(t,\tau) \right]$. The entries of $\mathbf{H}(t,\tau)$ consist of two components, i.e., the LOS component and the non-LOS (NLOS) component, and can be written as
\begin{align}
&h_{qp}(t,\tau)\nonumber\\
&=\underbrace{\sqrt{\frac{K(t)}{K(t)+1}}h^{\mathrm{LOS}}_{qp}(t)\delta\left(\tau-\tau^{\mathrm{LOS}}(t)\right)}_\mathrm{LOS}\nonumber\\
&+\underbrace{\sqrt{\frac{1}{K(t)+1}}\sum\limits_{n=1}^{N(t)}\sum\limits_{m_n=1}^{M_n(t)}h_{qp,n,m_n}(t)\delta\left(\tau-\tau_n(t)-\tau_{m_n}(t)\right)}_{\mathrm{NLOS}}.
\label{equ_cir_general}
\end{align}
In (\ref{equ_cir_general}), $K(t)$ is the Rician factor, $N(t)$ is the time variant number of clusters, $M_n(t)$ is the number of rays within $\mathrm{Cluster}_n$, $\tau_n(t)$ is the delay of $\mathrm{Cluster}_n$, and $\tau_{m_n}(t)$ is the relative delay of the $m_n$th ray in $\mathrm{Cluster}_n$. It is important to mention that all the parameters of the proposed 5G GBSM are time-variant, which has the capability to model the time-evolution and high mobility features of channels and is essentially a non-stationary channel model. To simplify the model, we assume that the Rician factor and relative delays are constants during the generation of channel coefficients, i.e., $K(t)=K$ and $\tau_{m_n}(t)=\tau_{m_n}$. These may not hold in certain scenarios such as the HST cutting scenario where the Rician factor is changing with time \cite{He13}. However, the cutting scenario is not frequently occurring in 5G scenarios and it is of high complexity. Also, the number of rays within a cluster is assumed to follow a Poisson distribution $\mathrm{Pois}\left(\tilde{\lambda} \right)$ \cite{Akdeniz14}, i.e., $M_n(t)=M_n=\max\left\lbrace \mathrm{Pois}\left(\tilde{\lambda} \right),1\right\rbrace$, where $\tilde{\lambda}$ is both the mean and variance of $M_n$ and $\max\left\lbrace \cdot \right\rbrace$ calculates the maximum value. Each ray within a cluster has its own complex gain and delay to support mmWave channels in the proposed 5G GBSM.

For the LOS component, if polarized antenna arrays are assumed at both the receiver and transmitter sides, the complex channel gain $h^{\mathrm{LOS}}_{qp}(t)$ is presented as (\ref{equ_LOS_component}),
\begin{figure*}[!t]
% ensure that we have normalsize text
\normalsize
% Store the current equation number.
% Set the equation number to one less than the one
% desired for the first equation here.
% The value here will have to changed if equations
% are added or removed prior to the place these
% equations are referenced in the main text.
\begin{align}
h^{\mathrm{LOS}}_{qp}(t)=&\begin{bmatrix}
F_{p,\mathrm{V}}^T(\mathbf{D}_{qp}^{\mathrm{LOS}}(t),\mathbf{A}_p^T(t))
\\
F_{p,\mathrm{H}}^T(\mathbf{D}_{qp}^{\mathrm{LOS}}(t),\mathbf{A}_p^T(t))
\end{bmatrix}^{\mathrm{T}}
\begin{bmatrix}
 e^{j\Phi_{\mathrm{LOS}}}& 0\\
0& -e^{j\Phi_{\mathrm{LOS}}}
\end{bmatrix}
\begin{bmatrix}
F_{q,\mathrm{V}}^R(\mathbf{D}_{qp}^{\mathrm{LOS}}(t),\mathbf{A}_q^R(t))
\\
F_{q,\mathrm{H}}^R(\mathbf{D}_{qp}^{\mathrm{LOS}}(t),\mathbf{A}_q^R(t))
\end{bmatrix}e^{j2\pi f_{qp}^\mathrm{LOS} (t)t}
\label{equ_LOS_component}
\end{align}
% Restore the current equation number.
% IEEE uses as a separator
\hrulefill
% The spacer can be tweaked to stop underfull vboxes.
\vspace*{4pt}
\end{figure*}
where $\Phi_{\mathrm{LOS}}$ is uniformly distributed within $(0,2\pi]$ \cite{36873}. The superscripts V and H denote vertical polarization and horizontal polarization, respectively. Functions $F^T(\mathbf{a},\mathbf{b})$ and $F^R(\mathbf{a},\mathbf{b})$ are antenna patterns with input vectors $\mathbf{a}$ and $\mathbf{b}$ in the GCS. The input vectors $\mathbf{a}$ and $\mathbf{b}$ need to be transformed into the LCS to obtain the antenna patterns. Detailed calculations are presented in Appendix~\ref{Appendix_Antenna_Pattern} \cite{36873} and the antenna pattern functions can be modified according to practical antenna settings. For large bandwidth support, the antenna patterns can be frequency dependent. In this case, the antenna patterns can be implemented accordingly in this model as well. The Doppler frequency $f_{qp}^{\mathrm{LOS}} (t)$ between $\mathrm{Ant}_q^{R}$ and $\mathrm{Ant}_p^{T}$ of the LOS component is expressed as $f_{qp}^{\mathrm{LOS}} (t)=\frac{1}{\lambda}\frac{\left\langle \mathbf{D}_{qp}^{\mathrm{LOS}}(t),\mathbf{v}^R-\mathbf{v}^T \right\rangle}{\left \| \mathbf{D}_{qp}^{\mathrm{LOS}}(t) \right \|}$
%\begin{align}
%f_{qp}^{\mathrm{LOS}} (t)=\frac{1}{\lambda}\frac{\left\langle \mathbf{D}_{qp}^{\mathrm{LOS}}(t),\mathbf{v}^R-\mathbf{v}^T \right\rangle}{\left \| \mathbf{D}_{qp}^{\mathrm{LOS}}(t) \right \|}
%\end{align}
where $\left\langle \cdot,\cdot \right\rangle$ is the inner product operator, $\|\cdot\|$ calculates the Frobenius norm, and $\lambda$ is the wavelength with respect to the central carrier frequency. Given the speed of light $c$, the delay $\tau^{\mathrm{LOS}}(t)$ of the LOS component is computed as $\tau^{\mathrm{LOS}}(t)=\left \| \mathbf{D}(t) \right \|/c$.
%\begin{align}
%\Phi_{qp}^{\mathrm{LOS}}(t)=\Phi_0+\frac{2\pi}{\lambda}\left \| \mathbf{D}_{qp}^{\mathrm{LOS}}(t) \right \|
%\end{align}
%\begin{align}
%\tau^{\mathrm{LOS}}(t)=\left \| \mathbf{D}(t) \right \|/c.
%\end{align}

For NLOS components, if $\mathrm{Cluster}_n$ is observable to $\mathrm{Ant}_{q}^{R}$ and $\mathrm{Ant}_{p}^{T}$, i.e., $\mathrm{Cluster}_n\in S_{qp}(t)$, the complex channel gain is expressed as (\ref{equ_channel_gain_NLOS}),
\begin{figure*}[!t]
% ensure that we have normalsize text
\normalsize
% Store the current equation number.
% Set the equation number to one less than the one
% desired for the first equation here.
% The value here will have to changed if equations
% are added or removed prior to the place these
% equations are referenced in the main text.
\begin{align}
h_{qp,n,m_n}(t)=&\begin{bmatrix}
F_{p,\mathrm{V}}^T(\mathbf{D}_{n,m_n}^T(t),\mathbf{A}_p^T(t))
\\
F_{p,\mathrm{H}}^T(\mathbf{D}_{n,m_n}^T(t),\mathbf{A}_p^T(t))
\end{bmatrix}^{\mathrm{T}}
\begin{bmatrix}
 e^{j\Phi_{n,m_n}^{\mathrm{VV}}}& \sqrt{\kappa}e^{j\Phi_{n,m_n}^{\mathrm{VH}}}\\
 \sqrt{\kappa}e^{j\Phi_{n,m_n}^{\mathrm{HV}}}& e^{j\Phi_{n,m_n}^{\mathrm{HH}}}
\end{bmatrix}
\begin{bmatrix}
F_{q,\mathrm{V}}^R(\mathbf{D}_{n,m_n}^R(t),\mathbf{A}_q^R(t))
\\
F_{q,\mathrm{H}}^R(\mathbf{D}_{n,m_n}^R(t),\mathbf{A}_q^R(t))
\end{bmatrix}\times\nonumber\\
&\sqrt{P_{n,m_n}(t)}e^{j2\pi f_{qn,m_n}^R (t)t}e^{j2\pi f_{pn,m_n}^T (t)t}
\label{equ_channel_gain_NLOS}
\end{align}
% Restore the current equation number.
% IEEE uses as a separator
\hrulefill
% The spacer can be tweaked to stop underfull vboxes.
\vspace*{4pt}
\end{figure*}
where $\kappa$ is the cross polarization power ratio and $P_{n,m_n}$ is the normalized mean power of the $m_n$th ray in $\mathrm{Cluster}_n$. The normalized mean power of $\mathrm{Cluster}_n$ can be calculated as $P_{n}=\sum\limits_{m_n} P_{n,m_n}$. Random phases $\Phi_{n,m_n}^{\mathrm{VV}},\Phi_{n,m_n}^{\mathrm{VH}},\Phi_{n,m_n}^{\mathrm{HV}},\Phi_{n,m_n}^{\mathrm{HH}}$ are uniformly distributed within $(0,2\pi]$ \cite{winner}.

Conversely, if $\mathrm{Cluster}_n$ is not observable, i.e., $\mathrm{Cluster}_n\notin S_{qp}(t)$, the complex channel gain
$h_{qp,n,m_n}(t)=0$.
%\begin{align}
%h_{qp,n,m_n}(t)=0.
%\end{align}
Accordingly, the Doppler frequencies at the receiver and transmitter are calculated as
%\begin{align}
%h_{qp,n,m_n}(t)=0
%\end{align}
\begin{align}
f_{qn,m_n}^R (t)=\frac{1}{\lambda}\frac{\left\langle \mathbf{D}_{qn,m_n}^R(t),\mathbf{v}^R-\mathbf{v}_n^R \right\rangle}{\left \| \mathbf{D}_{qn,m_n}^R(t) \right \|}
\end{align}
\begin{align}
f_{pn,m_n}^T (t)=\frac{1}{\lambda}\frac{\left\langle \mathbf{D}_{pn,m_n}^T(t),\mathbf{v}^T-\mathbf{v}_n^T \right\rangle}{\left \| \mathbf{D}_{pn,m_n}^T(t) \right \|}.
\end{align}
Moreover, the delay $\tau_n(t)$ of the NLOS component is computed as
\begin{align}
\tau_n(t)=\left[\left \| \mathbf{D}_n^R(t) \right \|+\left \| \mathbf{D}_n^T(t) \right \|\right]/c+\tilde{\tau}_n(t)
\end{align}
where $\tilde{\tau}_n(t)$ is an exponentially distributed random variable representing the virtual delay caused by the virtual link between the first and last bounces of $\mathrm{Cluster}_n$ in the scattering environment. The time evolution of virtual delays will be introduced in Section \ref{sec_survived_clusters}.
\subsection{Array-Time Cluster Evolution for the General 3D 5G GBSM}
\begin{figure}
\centering\includegraphics[width=3.5in]{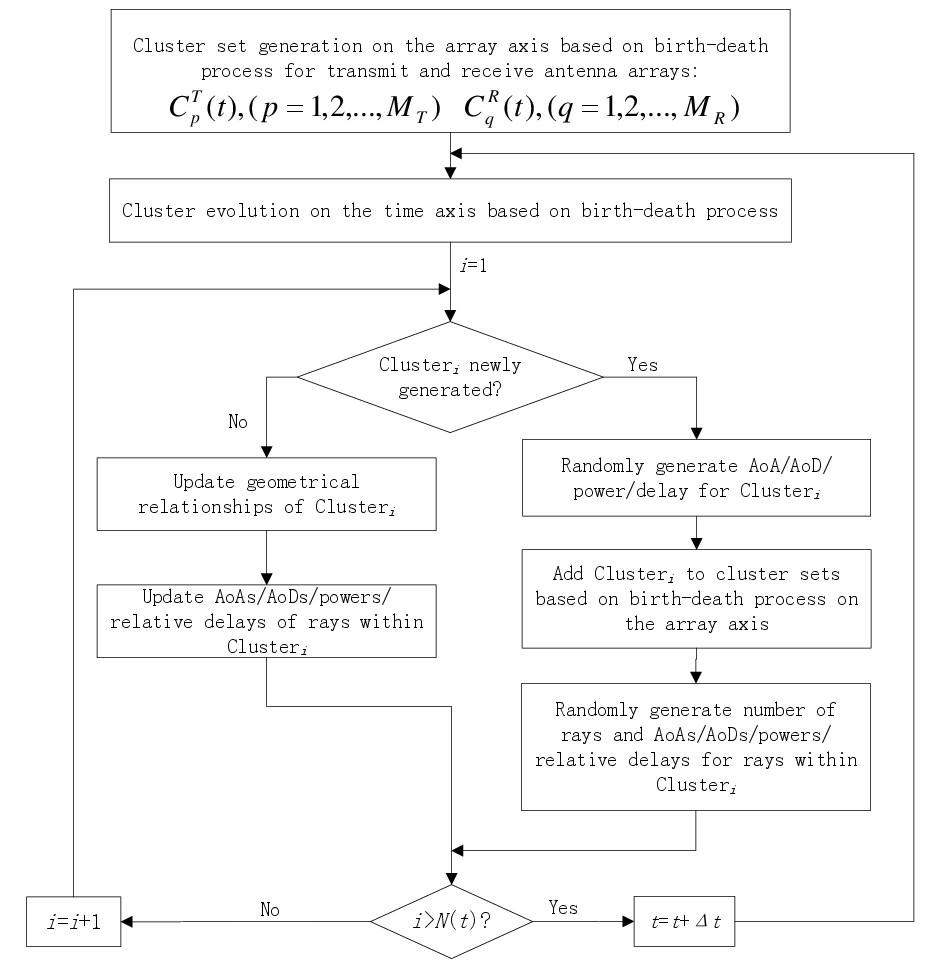}
\centering\caption{Flowchart of the array-time cluster evolution for the proposed general 3D non-stationary 5G GBSM.}
\label{fig_ArrayTimeBirthDeath4UnifiedFramework}
\end{figure}

The array-time cluster evolution for the proposed unified 5G GBSM framework is developed based on the birth-death process and the algorithm described in \cite{Wu14}. However, the proposed algorithm for the unified 5G GBSM framework in Fig. \ref{fig_ArrayTimeBirthDeath4UnifiedFramework} improves the one in \cite{Wu14} by including mean power evolution and updates of rays within clusters. Let us assume the generation (birth) and recombination (death) rates of clusters are $\lambda_G$ and $\lambda_R$, respectively. Then, the array-time cluster evolution for the unified 5G GBSM framework can be described as follows.

Step 1: An initial set of clusters are generated at time $t$. The generation procedure of initial clusters will be described in Section~\ref{sec_new_clusters}.

Step 2: At time $t+\Delta t$, cluster evolution on the time axis is operated. In principle, each cluster should have its own survival probability according to the relative motion. However, for simplicity, mean relative velocities of clusters will be used to calculate survival probabilities of clusters. Mean relative velocities $\Delta v^R$ and $\Delta v^T$ are characterized as $\Delta v^R=\mathrm{E}\left[ \left \| \mathbf{v}^R-\mathbf{v}^R_n \right \| \right]$ and $\Delta v^T=\mathrm{E}\left[ \left \| \mathbf{v}^T-\mathbf{v}^T_n \right \| \right]$, respectively.
As a result, the survival probability $P_{\mathrm{T}}(\Delta t)$ of a cluster after $\Delta t$ is computed as
\begin{equation}
P_{\mathrm{T}}(\Delta t)=e^{-\lambda_R \frac{P_F(\Delta v^R+\Delta v^T)\Delta t}{D^s_c} }
\end{equation}
where $P_F$ is the percentage of moving clusters and $D_c^s$ is a scenario-dependent coefficient describing space correlation. Typical values of $D_c^s$ such as $10$~m, $30$~m, $50$~m, and $100$~m can be chosen with the same order of correlation distances in \cite{38901}. The survivability of each cluster at time $t+\Delta t$ is determined by $P_{\mathrm{T}}(\Delta t)$. Geometrical relationships, delays, and mean powers of survived clusters will be updated according to Section \ref{sec_survived_clusters}. Meanwhile, a random number of new clusters are generated. This random number is generated according to a Poisson distribution with mean
\begin{equation}
\textrm{E}[N_{\textrm{new}}(t+\Delta t)] = {{\lambda _G } \over {\lambda _R }}(1 - P_{\mathrm{T}}(\Delta t) ).
\label{equ_time_evolution_avg_new}
\end{equation}
Rays and geometrical parameters will be assigned to these new clusters as described in Section \ref{sec_new_clusters}, which also presents the array axis evolution for clusters.

Step 3: When the array-time evolution is finished, the algorithm returns to Step 2 to enter the next time instant.

\subsection{Generation of New Clusters} \label{sec_new_clusters}
For a new cluster, say $\mathrm{Cluster}_n$, certain parameters such as the number of rays within the cluster, virtual delay, mean power, angular parameters, and relative delays of rays need to be assigned to this cluster. These parameters are randomly generated according to the distributions listed in Table \ref{tab_cluster_properties}. The virtual delays $\tilde{\tau}_{n}$ of clusters are assumed to be exponentially distributed, as in the WINNER II channel model \cite{winner}, and can be expressed by
\begin{align}
\tilde{\tau}_{n}=-r_{\tau}\sigma_{\tau}\cdot\mathrm{ln}u_n
\label{equ_virtual_delay}
\end{align}
where $u_n$ is uniformly distributed within $(0,1)$, $r_{\tau}$ is the delay scalar ($r_{\tau}=2.3$ for NLOS urban outdoor scenario and $r_{\tau}=2.4$ for NLOS indoor office scenario \cite{winner}), and $\sigma_{\tau}$ is a randomly generated delay spread ($\mathrm{E}\left[\log_{10}\sigma_{\tau}\right]=-6.63$ and $\mathrm{std}\left[\log_{10}\sigma_{\tau}\right]=0.32$ for NLOS urban outdoor scenario and $\mathrm{E}\left[\log_{10}\sigma_{\tau}\right]=-7.60$ and $\mathrm{std}\left[\log_{10}\sigma_{\tau}\right]=0.19$ for NLOS indoor office scenario \cite{winner}).
The mean powers $\tilde{P}'_{n}$ of clusters are generated as \cite{winner}
\begin{align}
\tilde{P}'_{n}=\exp\left(-\tilde{\tau}_{n}\frac{r_{\tau}-1}{r_{\tau}\sigma_{\tau}}\right)10^{-\frac{Z_{n}}{10}}
\label{equ_new_cluster_power}
\end{align}
where $Z_{n}$ follows a Gaussian distribution $\mathcal{N}(0,3)$ \cite{winner}. Unlike the virtual delay and mean power of clusters, which are generated as in the WINNER II channel model, generations of angular parameters, relative delays of rays, and mean power of rays are not following the WINNER II channel model. The angular parameters $\phi_{n}^{A}$, $\phi_{n}^{E}$, $\varphi_{n}^{A}$, and $\varphi_{n}^{E}$ of $\mathrm{Cluster}_n$ are assumed to obey wrapped Gaussian distributions. Angles of arrival (AoAs) of $\mathrm{Cluster}_n$ are generated as
\begin{align}
\phi_{n}^{A}=\mathrm{std}\left[\phi_{n}^{A} \right]Y^A_n+\psi_A^R
\label{equ_AAoA}
\end{align}
\begin{align}
\phi_{n}^{E}=\mathrm{std}\left[\phi_{n}^{E} \right]Y^E_n+\psi_E^R
\label{equ_EAoA}
\end{align}
where $Y^A_n,Y^E_n\sim \mathcal{N}(0,1)$, $\mathrm{std}\left[\phi_{n}^{A} \right]$ and $\mathrm{std}\left[\phi_{n}^{E} \right]$ are standard deviations of AoAs and need to be estimated. The parameter estimation procedure is introduced in Appendix \ref{Appendix_ParaEstimation}.
The mean power generation method of a cluster is extended to compute the mean power of rays within clusters as \cite{Akdeniz14}
%$\tilde{P}'_{n,m_n}=\exp\left(-\tau_{n,m_n}\frac{r_{\tau}-1}{r_{\tau}\mathrm{E}\left[\tau_{m_n} \right]}\right)10^{-\frac{Z_{n,m_n}}{10}}$
\begin{align}
\tilde{P}'_{n,m_n}=\exp\left(-\tau_{m_n}\frac{r_{\tau}-1}{\mathrm{E}\left[\tau_{m_n} \right]}\right)10^{-\frac{Z_{n,m_n}}{10}}
\end{align}
where $Z_{n,m_n}$ follows a Gaussian distribution $\mathcal{N}(0,3)$ \cite{winner}. The mean relative delay $\mathrm{E}\left[\tau_{m_n} \right]$ of rays of $\mathrm{Cluster}_n$ will be given in Section \ref{sec_numerical_analysis_section}. The mean power of rays is then scaled by the cluster power as $\tilde{P}_{n,m_n}=\tilde{P}'_{n}\frac{\tilde{P}'_{n,m_n}}{\sum\limits_{m_n}\tilde{P}'_{n,m_n}}$.
%\begin{align}
%\tilde{P}_{n,m_n}=\frac{\tilde{P}'_{n,m_n}}{\sum\limits_{m_n}\tilde{P}'_{n,m_n}}.
%\end{align}
Then, the angular parameters of $\mathrm{Cluster}_n$ via the $m_n$th ray can be calculated by adding the angular offset of the ray, i.e.,
\begin{align}
&\left[\phi_{n,m_n}^{A}\,\phi_{n,m_n}^{E}\,\varphi_{n,m_n}^{A}\,\varphi_{n,m_n}^{E}\right]^{\mathrm{T}}\nonumber\\
&=\left[\phi_{n}^{A}\,\phi_{n}^{E}\,\varphi_{n}^{A}\,\varphi_{n}^{E}\right]^{\mathrm{T}}+\left[\Delta\phi^{A}\,\Delta\phi^{E}\,\Delta\varphi^{A}\,\Delta\varphi^{E}\right]^{\mathrm{T}}
\end{align}
where $\Delta\phi^{A}$, $\Delta\phi^{E}$, $\Delta\varphi^{A}$, and $\Delta\varphi^{E}$ are angular offsets of the ray and are assumed to follow Laplace distributions \cite{winner} with zero mean and standard deviation of $1$ degree ($0.017$ radian) for simplicity. The standard deviation of angular offsets can be modified subject to measurements.
\begin{table}
\caption{Distributions of Parameters of Clusters.}
\center
    \begin{tabular}{|c|c|c|c|}
    \hline
    Parameters  & $M_n$ & $\tilde{\tau}_{n}$, $\tau_{m_n}$, $D_n^T, D_n^R$ & $\phi_{n}^{A}$, $\phi_{n}^{E}$,$\varphi_{n}^{A}$, $\varphi_{n}^{E}$  \\ \hline
        Distributions  & Poisson & Exponential & Wrapped Gaussian \\ \hline

    \end{tabular}
    \label{tab_cluster_properties}
\end{table}

Next, which antennas are able to observe the newly generated cluster should be determined. To avoid repeated description, only the receiver side is presented, the transmitter side follows the same procedure. First, the newly generated cluster is added to the cluster set of a randomly selected receive antenna $\mathrm{Ant}^{R}_{\tilde{q}}$. Second, we generate a 3D ball with radius $r\sim\exp \left(\frac{\lambda_R}{D^a_c} \right)$ where $D^a_c$ is the scenario-dependent coefficient normalizing antenna spacings. Typical values of $D^a_c$ such as $30$~m and $50$~m can be chosen with the same order of correlation distances in \cite{38901}. Third, we compute the distances between antennas $\mathrm{Ant}^{R}_{\tilde{q}}$ and $\mathrm{Ant}^{R}_{{q}}$ for all $q$. Then, we add the newly generated cluster to the cluster sets of antennas satisfying $\|\mathbf{A}^{R}_q-\mathbf{A}^{R}_{\tilde{q}} \|\leq r$. As a consequence, the probability that both $\mathrm{Ant}^{R}_{\tilde{q}}$ and $\mathrm{Ant}^{R}_{{q}}$ are able to observe this newly generated cluster will be $\exp \left(\frac{\lambda_R}{D^a_c}\|\mathbf{A}^{R}_q-\mathbf{A}^{R}_{\tilde{q}} \| \right)$. The psuedo codes of the new cluster generation algorithm are shown in Fig.~\ref{fig_algorithm}.
%
%The survival probability $P_{\mathrm{A}}$ of a cluster evolving from one antenna to an adjacent antenna can be computed as
%$P_{\mathrm{A}}=e^{-\lambda_R\frac{\delta_R}{D^a_c}},$
%%\begin{equation}
%%P_{\mathrm{A}}=e^{-\lambda_R\frac{\delta_R}{D^a_c}}
%%\end{equation}
% Then,
%
% The cluster evolves based on birth-death process to neighboring antennas of the selected receive antenna with survival probability $P_A$.

\begin{figure}
\hrulefill
\begin{algorithmic}[1]
\State Generate $M_n\sim\mathrm{Pois}\left(\tilde{\lambda} \right)$ rays for the cluster;
\State Generate virtual delay/mean power/distance for the cluster;
\State Generate AoAs/Angles of departure (AoDs)/relative delays/relative mean powers for the rays within the cluster;
\State Generate discrete $\tilde{q}\sim\mathrm{U}\left(1,M_R \right)$, generate $r\sim\exp \left(\frac{\lambda_R}{D^a_c} \right)$, and let $\tilde{i}=1$;
%\State Generate $r\sim\exp \left(\frac{\lambda_R}{D^a_c} \right)$;

\While {($\tilde{i}\leq M_R$)}
\If {($\|\mathbf{A}^{R}_{\tilde{i}}-\mathbf{A}^{R}_{\tilde{q}} \|\leq r$)}
    \State Add the cluster to $C^R_{\tilde{i}}(t)$;
    \EndIf
        \State $\tilde{i}=\tilde{i}+1$;

    %\State $\tilde{i}=\tilde{i}+1$;
    \EndWhile

%\While {($\tilde{u}\sim\mathrm{U}\left(0,1 \right)\leq P_\mathrm{A}$\& $\tilde{q}-\tilde{j}\geq 1$)}
%    \State Add the cluster to $C^R_{(\tilde{q}-\tilde{j})}(t)$; $\tilde{j}=\tilde{j}+1$;
%    %\State $\tilde{j}=\tilde{j}+1$;
%\EndWhile
\end{algorithmic}
\hrulefill
\caption{Psuedo codes for the new cluster generation algorithm.}
\label{fig_algorithm}
\end{figure}
\subsection{Evolution of Survived Clusters}\label{sec_survived_clusters}
In order to highlight time evolution of the proposed model, geometrical relationships, virtual delays, and mean powers of survived clusters need to be updated from $t$ to $t+\Delta t$. To begin with, antenna position vectors are updated as
\begin{align}
\mathbf{A}_q^R(t+\Delta t)=\mathbf{A}_q^R(t)+\mathbf{v}^R\Delta t
\end{align}
\begin{align}
\mathbf{A}_p^T(t+\Delta t)=\mathbf{A}_p^T(t)+\mathbf{v}^T\Delta t.
\end{align}
At the same time, distance vectors of clusters need to be adjusted as
\begin{align}
\mathbf{D}_n^R(t+\Delta t)=\mathbf{D}_n^R(t)+\mathbf{v}_n^R\Delta t
\end{align}
\begin{align}
\mathbf{D}_n^T(t+\Delta t)=\mathbf{D}_n^T(t)+\mathbf{v}_n^T\Delta t.
\end{align}
Other distance vectors in (\ref{equ_D_qp_LOS})--(\ref{equ_D_T_pn_m_n}) can be updated accordingly. Delays are updated as
\begin{align}
&\tau_n(t+\Delta t)\nonumber\\
&=\left[\left \| \mathbf{D}_n^R(t+\Delta t) \right \|+\left \| \mathbf{D}_n^T(t+\Delta t) \right \|\right]/c+\tilde{\tau}_n(t+\Delta t).
\end{align}
The random virtual delays $\tilde{\tau}_n(t+\Delta t)$ are modeled as
$\tilde{\tau}_n(t+\Delta t)=e^{-\frac{\Delta t}{\varsigma}}\tilde{\tau}_n(t)+(1-e^{-\frac{\Delta t}{\varsigma}})X$
where $X$ is a random variable independent to $\tilde{\tau}_{n}$ but identically distributed as $\tilde{\tau}_{n}$ and $\varsigma$ is a scenario-dependent parameter describing the coherence of virtual links. Typical values of the coherence of virtual links such as $5$~s, $7$~s, and $30$~s are chosen based on reasonable assumptions. Thus, the updated delay will carry information of the delay in the previous time instant.

Another important aspect is the evolution of cluster mean power. Constant cluster mean powers were assumed in \cite{Wu15} and \cite{Wu14}, which were not sufficient to characterize time evolution of the channel. Therefore, in this paper, with the assumption that the cluster mean powers satisfy the inverse square law, the time evolution of cluster mean power can be expressed as (derivations given in Appendix \ref{Appendix_Cluster_Power})
\begin{align}
\tilde{ P}_{n,m_n}(t+\Delta t)=\tilde{ P}_{n,m_n}(t)\frac{3\tau_n(t)-2\tau_n(t+\Delta t)+\tau_{m_n}}{\tau_n(t)+\tau_{m_n}}.
\label{equ_Pn_evolution}
\end{align}
The mean power terms $\tilde{ P}_{n,m_n}$ in the mean power evolution in (\ref{equ_Pn_evolution}) are not normalized. They need to be normalized such that $P_{n,m_n}=\tilde{ P}_{n,m_n}/\sum\limits_{n,m_n} \tilde{ P}_{n,m_n}$ before being plugged into (\ref{equ_channel_gain_NLOS}). To guarantee smooth power transitions when clusters appear or disappear, a simple linear power scaling is performed within 1 ms. That is to say, the power of a disappearing cluster will be scaled linearly from its instant power to zero within 1 ms, and the power of a newly generated cluster will be scaled from 0 to its power shown in (\ref{equ_new_cluster_power}). The choice of a 1 ms transition period is aligned with the length of one subframe in LTE \cite{36211}, which is easier for system-level simulations with the proposed channel model. Other non-linear transitions can be found in \cite{Jaeckel14}.

\subsection{Simplified Channel Models}\label{sec_adapt_scenarios}
The proposed general 3D non-stationary 5G GBSM can easily be reduced to various simplified channel models by adjusting certain model parameters.
\begin{enumerate}
\item By setting the numbers of antennas ($M_R$ and $M_T$) as relatively small numbers, the spherical wavefront effect and cluster evolution on the array axis will become insignificant. In this case, the general 5G massive MIMO channel model is reduced to a conventional (non-massive) MIMO channel model.
\item By setting the velocity of the transmitter $\mathbf{v}^T=\mathbf{0}$, the general 5G V2V channel model is reduced to a fixed-to-mobile (F2M) channel model.
\item By setting the relative delays of rays as $0$, i.e., $\tau_{m_n}=0$, rays within a cluster will become irresolvable in the delay domain. Consequently, the general 5G mmWave channel model is simplified as a SCM-like wideband channel model.
\item By setting all elevation angles as zero, i.e., $\psi_E^{R}=\psi_E^{T}=\phi_n^{E}(t)=\varphi_n^{E}(t)=0$, the impacts of elevation angles are ignored. Then, the 3D 5G channel model is simplified to 2D.
\end{enumerate}
By properly adjusting the parameters of the proposed general 5G channel model, various simplified channel models can be obtained, such as 3D wideband massive MIMO, 3D wideband HST conventional MIMO, 3D mmWave conventional MIMO, and 2D wideband V2V conventional MIMO channel models.
%\begin{figure}
%\centering\includegraphics[width=4in]{Unified_framework_adaptations.eps}
%\centering\caption{Diagram of combinations of scenarios from the proposed unified 5G GBSM framework.}
%\label{fig_Unified_framework_adaptations}
%\end{figure}
%%%%%%%%%%%{Wideband Twin Cluster MIMO Model Model ends%%%%%%%%%%%%%%

%%%%%%%%%%%Stochastic Property Analysis begins%%%%%%%%%%%%
\section{Statistical Property Analysis} \label{sec_StochasticPropertyAnalysis}
\subsection{Time-Variant Power Delay Profile (PDP)}
The time-variant PDP $\Lambda(t,\tau)$ of the channel can be expressed as \cite{MobileRadioChan}
\begin{align}
\Lambda(t,\tau)=\sum\limits_n^{N(t)}\sum\limits_{m_n}^{M_n}P_{n,m_n}(t)\delta\left(\tau-\tau_n(t)-\tau_{m_n}\right).
\label{equ_time_variant_pdp}
\end{align}
It should be noted that the PDP is from all observable clusters at both the transmit and receive arrays. The time-variant properties of PDP are caused by the time-dependent mean powers and delays of rays. These are related to the geometrical relationship updates of the scattering environment.

\subsection{Stationary Interval}
The stationary interval is utilized in \cite{Chen12} to measure the estimated period within which the channel amplitude response can be regarded as stationary. It can be used to determine the frequency of channel estimation in HST communications. The definition of the stationary interval is the maximum time length within which the autocorrelation function (ACF) of the PDP exceeds the $80\%$ threshold \cite{Chen12}. It should be noticed that the $80\%$ threshold in \cite{Chen12} is empirical and can be adjusted according to requirements.  Also, the definition of stationary interval in \cite{Chen12} would fail to work if the ACF of the PDP is not monotonically decreasing or has multiple crossing points at the threshold. Therefore, an improved definition of the stationary interval $I(t)$ at time $t$ is proposed as
$I(t)=\mathrm{inf}\left\lbrace \Delta t|_{R_{\Lambda} (t,\Delta t)\leq 0.8} \right\rbrace$
%\begin{align}
%I(t)=\mathrm{inf}\left\lbrace \Delta t|_{R_{\Lambda} (t,\Delta t)\leq 0.8} \right\rbrace
%\end{align}
where $\mathrm{inf}\left\lbrace \cdot \right\rbrace$ calculates the infimum of a function and $R_{\Lambda (t,\Delta t)}$ is the normalized ACF of the PDP defined by \cite{Chen12}
\begin{align}
R_{\Lambda} (t,\Delta t)=\frac{\int \Lambda(t,\tau)\Lambda(t+\Delta t,\tau)d\tau}{\max\left\lbrace \int \Lambda^2(t,\tau)d\tau,\int \Lambda^2(t+\Delta t,\tau)d\tau \right\rbrace}.
\end{align}

\subsection{Time-Variant Transfer Function}
The time-variant transfer function $H_{qp}(\xi,t)$ is the Fourier transform of the channel impulse response with respect to delay, which can be expressed as (\ref{equ_time_variant_transfer_function}),
\begin{figure*}[!t]
% ensure that we have normalsize text
\normalsize
% Store the current equation number.
% Set the equation number to one less than the one
% desired for the first equation here.
% The value here will have to changed if equations
% are added or removed prior to the place these
% equations are referenced in the main text.
\begin{align}
H_{qp}(\xi,t)&=\int\limits_{-\infty}^{\infty}h_{qp}(t,\tau)e^{-j2\pi\xi\tau}d\tau={\sqrt{\frac{K}{K+1}}h^{\mathrm{LOS}}_{qp}(t)e^{-j2\pi\xi\tau^{\mathrm{LOS}}(t)}}+{\sqrt{\frac{1}{K+1}}\sum\limits_{n=1}^{N(t)}\sum\limits_{m_n=1}^{M_n(t)}h_{qp,n,m_n}(t)e^{-j2\pi\xi\left[\tau_n(t)+\tau_{m_n} \right]}}
\label{equ_time_variant_transfer_function}
\end{align}
% Restore the current equation number.
% IEEE uses as a separator
\hrulefill
% The spacer can be tweaked to stop underfull vboxes.
\vspace*{4pt}
\end{figure*}
where $\xi$ is frequency \cite{MobileRadioChan}.

\subsection{Space-Time-Frequency Correlation Function}

To study the correlation properties, the space-time-frequency correlation function (STFCF) $\rho _{qp,q'p'} (\|\mathbf{A}^{T}_p-\mathbf{A}^{T}_{{p'}} \| ,\|\mathbf{A}^{R}_q-\mathbf{A}^{R}_{{q'}} \|,\Delta\xi,\Delta t;\xi,t )$ can be calculated as  \cite{MobileRadioChan}
\begin{align}
&\rho _{qp,q'p'} (\|\mathbf{A}^{T}_p-\mathbf{A}^{T}_{{p'}} \| ,\|\mathbf{A}^{R}_q-\mathbf{A}^{R}_{{q'}} \|,\Delta\xi,\Delta t;\xi,t )\nonumber\\
&=\mathrm{E}\left[H^*_{qp}(\xi,t)H_{q'p'}(\xi+\Delta\xi,t+\Delta t)\right].
\label{equ_STFCF_general}
\end{align}
The LOS component is computed based on the relative position between the transmitter and receiver. The parameters of the NLOS components are randomly generated. Therefore, the LOS and NLOS components are assumed to be uncorrelated for simplicity. Then, (\ref{equ_STFCF_general}) can be written as the sum of the correlation of the LOS component and the correlation of the NLOS components, i.e.,
\begin{align}
\rho _{qp,q'p'} &(\|\mathbf{A}^{T}_p-\mathbf{A}^{T}_{{p'}} \| ,\|\mathbf{A}^{R}_q-\mathbf{A}^{R}_{{q'}} \|,\Delta\xi,\Delta t;\xi,t )\nonumber\\
&=\rho^{\mathrm{LOS}} _{qp,q'p'} (\|\mathbf{A}^{T}_p-\mathbf{A}^{T}_{{p'}} \| ,\|\mathbf{A}^{R}_q-\mathbf{A}^{R}_{{q'}} \|,\Delta\xi,\Delta t;\xi,t )\nonumber\\
&+\rho^{\mathrm{NLOS}} _{qp,q'p'} (\|\mathbf{A}^{T}_p-\mathbf{A}^{T}_{{p'}} \| ,\|\mathbf{A}^{R}_q-\mathbf{A}^{R}_{{q'}} \|,\Delta\xi,\Delta t;\xi,t ).
\label{equ_STFCF_LOS_NLOS}
\end{align}
The correlation of the LOS component is calculated as
\begin{align}
\rho^{\mathrm{LOS}} _{qp,q'p'} (& \|\mathbf{A}^{T}_p-\mathbf{A}^{T}_{{p'}} \| ,\|\mathbf{A}^{R}_q-\mathbf{A}^{R}_{{q'}} \|,\Delta\xi,\Delta t;\xi,t )\nonumber\\
&=\frac{K}{K+1}h^{\mathrm{LOS}*}_{qp}(t)h^{\mathrm{LOS}}_{q'p'}(t+\Delta t)e^{j2\pi\sigma_1 }
\label{equ_STFCF_LOS}
\end{align}
with $\sigma_1=\xi\left[ \tau^{\mathrm{LOS}}(t)-\tau^{\mathrm{LOS}}(t+\Delta t) \right]+\Delta\xi\tau^{\mathrm{LOS}}(t+\Delta t).$
%\begin{align}
%\sigma_1=\xi\left[ \tau^{\mathrm{LOS}}(t)-\tau^{\mathrm{LOS}}(t+\Delta t) \right]+\Delta\xi\tau^{\mathrm{LOS}}(t+\Delta t).
%\end{align}
Similarly, the correlation of the NLOS components is calculated as (\ref{equ_STFCF_NLOS}),
\begin{figure*}[!t]
% ensure that we have normalsize text
\normalsize
% Store the current equation number.
% Set the equation number to one less than the one
% desired for the first equation here.
% The value here will have to changed if equations
% are added or removed prior to the place these
% equations are referenced in the main text.
\begin{align}
\rho^{\mathrm{NLOS}} _{qp,q'p'} (\|\mathbf{A}^{T}_p-\mathbf{A}^{T}_{{p'}} \| ,\|\mathbf{A}^{R}_q-\mathbf{A}^{R}_{{q'}} \|,\Delta\xi,\Delta t;\xi,t )=\frac{1}{K+1}\mathrm{E}\left[\sum\limits_{n=1}^{N(t)}\sum\limits_{n'=1}^{N(t+\Delta t)}\sum\limits_{m_n=1}^{M_n}\sum\limits_{m_{n'}=1}^{M_{n'}}h_{qp,n,m_n}^{*}(t)h_{q'p',n',m_{n'}}(t+\Delta t)e^{j2\pi \sigma_2} \right]
\label{equ_STFCF_NLOS}
\end{align}
% Restore the current equation number.
% IEEE uses as a separator
\hrulefill
% The spacer can be tweaked to stop underfull vboxes.
\vspace*{4pt}
\end{figure*}
with $\sigma_2=\xi(\tau_n(t)+\tau_{m_n}-\tau_{n'}(t+\Delta t)-\tau_{m_{n'}})-\Delta\xi(\tau_{n'}(t+\Delta t)+\tau_{m_{n'}}).$
%\begin{align}
%\sigma_2=\xi(\tau_n(t)+\tau_{m_n}-\tau_{n'}(t+\Delta t)-\tau_{m_{n'}})-\Delta\xi(\tau_{n'}(t+\Delta t)+\tau_{m_{n'}}).
%\end{align}
Because cluster evolution is considered in the proposed general 5G GBSM, the mean number of survived cluster shared by $h_{qp,n,m_n}(t)$ and $h_{q'p',n',m_{n'}}(t+\Delta t)$ can be calculated as
\begin{align}
\mathrm{E} & \left\lbrace  \textrm{card}\left(S_{qp}(t)\bigcap S_{q'p'}(t+\Delta t)\right)\right\rbrace\nonumber\\
&=P_{\mathrm{survival}}\mathrm{E}\left\lbrace\textrm{card}\left(S_{qp}(t)\right)\right\rbrace
\end{align}
where $P_{\mathrm{survival}}$ is the cluster survival probability when a cluster evolves from $\mathrm{Ant}_p^{T}$, $\mathrm{Ant}_q^{R}$, and $t$ to $\mathrm{Ant}_{p'}^{T}$, $\mathrm{Ant}_{q'}^{R}$, and $t+\Delta t$, respectively, i.e.,
\begin{equation}
P_{\mathrm{survival}}=e^{-\lambda_R\left[\frac{\|\mathbf{A}^{T}_p-\mathbf{A}^{T}_{{p'}} \| +\|\mathbf{A}^{R}_q-\mathbf{A}^{R}_{{q'}} \|}{D^a_c}+\frac{P_F(\Delta v^R+\Delta v^T)\Delta t}{D^s_c} \right]}.
\end{equation}
Those newly generated clusters from $h_{qp,n,m_n}(t)$ to $h_{q'p',n',m_{n'}}(t+\Delta t)$ are independent to the survived clusters. Therefore, they do not contribute to the correlation coefficient. Then, the STFCF for NLOS components in (\ref{equ_STFCF_NLOS}) reduces to (\ref{equ_reduced_rho_NLOS}).
\begin{figure*}[!t]
% ensure that we have normalsize text
\normalsize
% Store the current equation number.
% Set the equation number to one less than the one
% desired for the first equation here.
% The value here will have to changed if equations
% are added or removed prior to the place these
% equations are referenced in the main text.
\begin{align}
&\rho^{\mathrm{NLOS}} _{qp,q'p'} (\delta _T ,\delta _R,\Delta\xi,\Delta t;\xi,t )=\frac{P_{\mathrm{survival}}}{K+1}\mathrm{E}\left[\sum\limits_{n=1}^{N(t)}\sum\limits_{n'=1}^{N(t)}\sum\limits_{m_n=1}^{M_n}\sum\limits_{m_{n'}=1}^{M_{n'}}h_{qp,n,m_n}^{*}(t)h_{q'p',n',m_{n'}}(t+\Delta t)e^{j2\pi \sigma_2} \right]
\label{equ_reduced_rho_NLOS}
\end{align}
% Restore the current equation number.
% IEEE uses as a separator
\hrulefill
% The spacer can be tweaked to stop underfull vboxes.
\vspace*{4pt}
\end{figure*}

%With the uncorrelated scattering (US) assumption, i.e., clusters are mutually independent
%\begin{equation}
%\mathrm{E}\left[h^*_{qp,n,m_n}(t) h_{qp,n',m_{n'}} (t+\Delta t) \right]=0
%\end{equation}
%for $n'\neq n$.
As the dimension of the STFCF is high, it is difficult to present the STFCF visually. However, by setting $\Delta \xi=0$, $q=q'$, and $p=p'$, the STFCF is reduced to the time-variant ACF. By setting $\Delta t=0$, $\Delta \xi=0$ and $p=p'$ ($q=q'$), the STFCF is reduced to the receive (transmit) space cross-correlation function (CCF).  Similarly, by setting $q=q'$, $p=p'$, and $\Delta t=0$, the STFCF is reduced to the time-variant frequency correlation function (FCF).
%By setting $q=q'$, $p=p'$, and $\Delta \xi=0$, the STFCF is reduced to the time-variant time correlation function (TCF).
%\subsection{Doppler Power Spectral Density}
%The Doppler Power Spectral Density (PSD) $S(\xi_\mathrm{D};\xi,t)$ at time $t$ and frequency $\xi$ can be computed as the Fourier transform of TCF in terms of $\Delta t$  \cite{MobileRadioChan}
%\begin{align}
%S(\xi_\mathrm{D};\xi,t)=\int_{-\infty}^{+\infty}\rho^{\mathrm{NLOS}} _{qp,q'p'} (0 ,0,0,\Delta t;\xi,t )e^{-j\xi_{\mathrm{D}}\Delta t}d(\Delta t)
%\label{equ_Doppler_PSD}
%\end{align}
%where $\xi_{\mathrm{D}}$ is the Doppler frequency.

%%%%%%%%%%%Stochastic Property Analysis ends%%%%%%%%%%%%%%

%%%%%%%%%%%Numerical Analysis begins%%%%%%%%%%%%
\section{Results and Analysis} \label{sec_numerical_analysis_section}
In the simulations, we assume that the generation rate $\lambda_\mathrm{G}=80/$m \cite[Table~I]{Zwick00} and the recombination rate $\lambda_\mathrm{R}=4/$m, so that the mean number of clusters is fixed as $20$ \cite{winner}. The percentage of moving clusters is $P_F=0.3$ \cite{Zwick00}. The parameter estimation of the proposed 5G small-scale fading channel model is based on the minimum mean square error (MMSE) criterion, i.e., $\epsilon=|\hat{\mathsf{f}}-\mathsf{f}(\mathsf{P})|^2$, where $\hat{\mathsf{f}}$ is the measured statistical property such as space CCF and complementary cumulative distribution function (CCDF) of coherence bandwidth, $\mathsf{f}$ is the statistical property of the channel model, and $\mathsf{P}$ is the parameter set of the channel model. Note that the parameter set $\mathsf{P}$ for a specific statistical property is not necessary to include all parameters of the channel model, since only a small number of parameters will significantly affect this statistical property. Besides the parameter set $\mathsf{P}$, the rest parameters are generated randomly according to the WINNER II channel model~\cite{winner}. Based on the MMSE criterion, an exhaustive search of the parameter set $\mathsf{P}$ will be performed using an optimization procedure until $\epsilon$ reaches a certain threshold and then the parameter set will be obtained. Details of the parameter estimation procedure are introduced in Appendix \ref{Appendix_ParaEstimation}. The cluster parameters finally obtained for simulations are listed in Table \ref{tab_simulation_parameters}. Their distributions can be referred to Table \ref{tab_cluster_properties}.

\begin{table*}[!htb]
\caption{Simulation parameters for different simplified 5G channel models.}
\center
\footnotesize
\begin{tabular}{|c|c|c|c|c|c|}
\hline

\multicolumn{3}{|c|}{3D wideband massive MIMO} & \multicolumn{3}{c|}{3D wideband HST conventional MIMO} \\ \hline
   Parameters    &  Mean     & Standard deviation      &   Parameters    &  Mean     &  Standard deviation     \\ \hline
   $M_n$    &   20    &    0   &   $M_n$    &  20     &   0    \\ \hline
   $\tau_n$    &    930 ns   &   930 ns    &  $\tau_n$     &   930 ns    &   930 ns     \\ \hline
%   $\tilde{P}_n$    &       &       &   $\tilde{P}_n$    &       &       \\ \hline
   $\tau_{m_n}$    &   0 ns    &   0 ns    &   $\tau_{m_n}$    &   0 ns    &    0 ns   \\ \hline

    $\left(D^R_n, D^T_n \right) $   &  $\left(  25, 30 \right) $  m   &  $\left(  15, 10\right) $  m   &    $\left(D^R_n,D^T_n \right) $   &    $\left(25, 30\right) $ m   &   $\left(15, 10\right) $ m      \\ \hline
     %  &   30 m    &   10 m    &        &  30 m    &   10 m \\ \hline
    $\phi_{n}^{A}$   &   0.78 rad    &   1.15 rad    &   $\phi_{n}^{A}$    &  0.78 rad     &  0.90 rad \\ \hline
    $\phi_{n}^{E}$   &    0.78 rad   &   0.18 rad    &   $\phi_{n}^{E}$    &    0.78 rad   &   0.18 rad      \\ \hline
   $\varphi_{n}^{A}$    &   1.05 rad    &   0.54 rad    &$\varphi_{n}^{A}$       &  1.05 rad    &   0.54 rad \\ \hline
   $\varphi_{n}^{E}$    &   0.78 rad    &  0.11 rad     &$\varphi_{n}^{E}$       &      0.78 rad    &  0.11 rad       \\ \hline

   $D^a_c$   &    30 m   &   0 m   &   $D^a_c$    &    50 m   &   0 m     \\ \hline
   $D^s_c$    &   100 m    &   0 m    &$D^s_c$       &  100 m    &   0 m \\ \hline
   $\varsigma$    &   30 s    &  0 s     &$\varsigma$       &      7 s    &  0 s       \\ \hline
   $\left(\psi^R_A, \psi^R_E \right)$   &    $\left(\frac{\pi}{4}, \frac{\pi}{4} \right)$ rad   &   $(0,0)$    &   $\left(\psi^R_A, \psi^R_E \right)$  rad  &  $\left(\frac{\pi}{4}, \frac{\pi}{4} \right)$ & $(0,0)$     \\ \hline

    $\left(\psi^T_A, \psi^T_E \right)$   &    $\left(\frac{\pi}{3}, \frac{\pi}{4} \right)$ rad  &   $(0,0)$    &   $\left(\psi^T_A, \psi^T_E \right)$  rad  &    $\left(\frac{\pi}{3}, \frac{\pi}{4} \right)$    &   $(0,0)$       \\ \hline

\multicolumn{3}{|c|}{2D wideband V2V conventional MIMO} & \multicolumn{3}{c|}{3D mmWave conventional MIMO} \\ \hline
      Parameters    &  Mean     & Standard deviation      &   Parameters    &  Mean     &  Standard deviation     \\ \hline
   $M_n$    &   20    &    0   &   $M_n$    &   15    &   3.87    \\ \hline
   $\tau_n$    &   930 ns    &   930 ns    &  $\tau_n$     &   305 ns    &  305 ns     \\ \hline
%   $\tilde{P}_n$    &       &       &   $\tilde{P}_n$    &       &       \\ \hline
   $\tau_{m_n}$    &   0 ns    &  0 ns     &   $\tau_{m_n}$    &   3 ns    &   3 ns    \\ \hline

    $\left(D^R_n, D^T_n \right) $   &  $\left(  25, 30 \right) $  m   &  $\left(  15, 10\right) $  m   &    $\left(D^R_n,D^T_n \right) $   &    $\left(5, 5\right) $ m   &   $\left(3, 3\right) $ m      \\ \hline
  % $D_n^R$   &      25 m   &   15 m   &   $D^R_n$    &    5 m   &   3 m    \\ \hline
    $\phi_{n}^{A}$   &       0.78 rad    &   0.91 rad       &   $\phi_{n}^{A}$    &        0.78 rad    &   0.91 rad       \\ \hline
    $\phi_{n}^{E}$   &       0 rad   &   0 rad       &   $\phi_{n}^{E}$    &       0.78 rad   &   0.18 rad       \\ \hline
   $\varphi_{n}^{A}$    &       1.04 rad    &   0.53 rad       &$\varphi_{n}^{A}$       &       1.04 rad    &   0.53 rad \\ \hline
   $\varphi_{n}^{E}$    &       0 rad    &  0 rad       &$\varphi_{n}^{E}$       &       0.78 rad    &  0.11 rad       \\ \hline
      $D^a_c$   &    30 m   &   0 m   &   $D^a_c$    &    30 m   &   0 m     \\ \hline
   $D^s_c$    &   10 m    &   0 m    &$D^s_c$       &  100 m    &   0 m \\ \hline
   $\varsigma$    &   5 s    &  0 s     &$\varsigma$       &      7 s    &  0 s       \\ \hline
 $\left(\psi^R_A, \psi^R_E \right)$   &    $\left(\frac{\pi}{4}, \frac{\pi}{4} \right)$ rad  &   $(0,0)$    &   $\left(\psi^R_A, \psi^R_E \right)$    &  $\left(\frac{\pi}{4}, \frac{\pi}{4} \right)$ rad & $(0,0)$     \\ \hline

    $\left(\psi^T_A, \psi^T_E \right)$   &    $\left(\frac{\pi}{3}, \frac{\pi}{4} \right)$  rad &   $(0,0)$    &   $\left(\psi^T_A, \psi^T_E \right)$    &    $\left(\frac{\pi}{3}, \frac{\pi}{4} \right)$  rad  &   $(0,0)$       \\ \hline
\end{tabular}
\label{tab_simulation_parameters}
\end{table*}

\begin{figure}
\centering\includegraphics[width=3.3in]{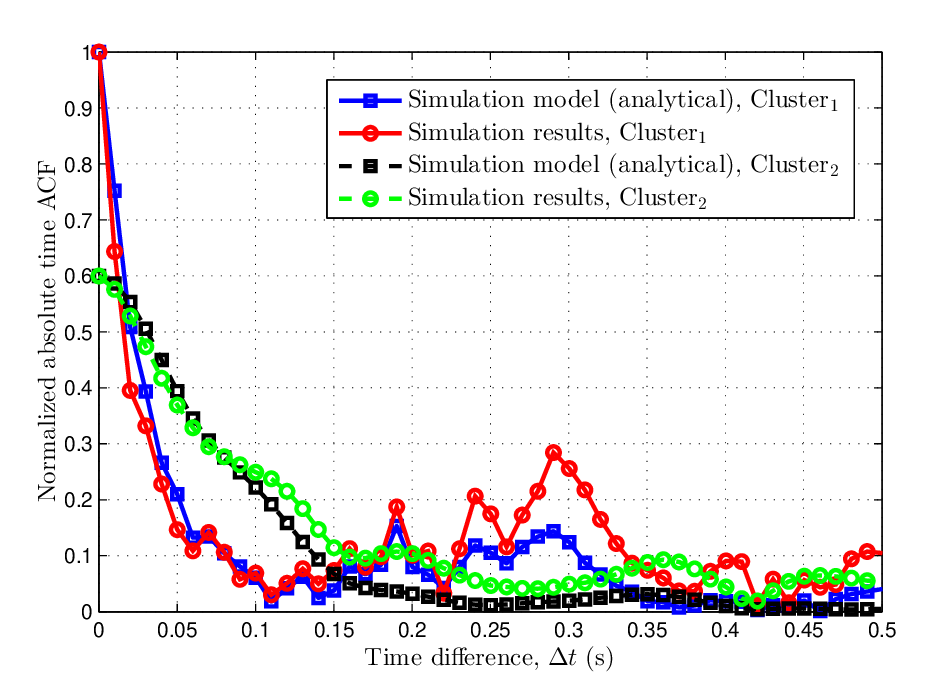}
\centering\caption{Comparison between the normalized analytical and simulated ACFs of $\mathrm{Cluster}_1$ and $\mathrm{Cluster}_2$ of the  wideband conventional MIMO channel model ($f_c=2$~GHz, $\|\mathbf{D}\|=200$~m, $M_R=M_T=2$, $\Delta v^R=\Delta v^T=0$~m/s, $|\mathbf{v}^T|=0$,$|\mathbf{v}^R|=5$~m/s, $D_c^a=50$~m, $D_c^s=100$~m, $M_1=M_2=81$, $\varsigma=7$~s, NLOS).}
\label{fig_ACF_simModel_sim}
\end{figure}

% Simulation model vs simulation
To validate the correctness of the proposed general 5G GBSM, the analytical ACFs of the simulation model and corresponding simulation results of the wideband conventional MIMO channel model for both $\mathrm{Cluster}_1$ and $\mathrm{Cluster}_2$ are compared in Fig. \ref{fig_ACF_simModel_sim}. Half-wavelength linear arrays are assumed at both the transmitter and receiver. It should be noted that these ACFs are normalized with respect to $\mathrm{Cluster}_1$. It can be seen from Fig. \ref{fig_ACF_simModel_sim} that the simulated ACFs align well with the analytical ACFs of the simulation model for both $\mathrm{Cluster}_1$ and $\mathrm{Cluster}_2$.

\begin{figure}
\centering\includegraphics[width=3.3in]{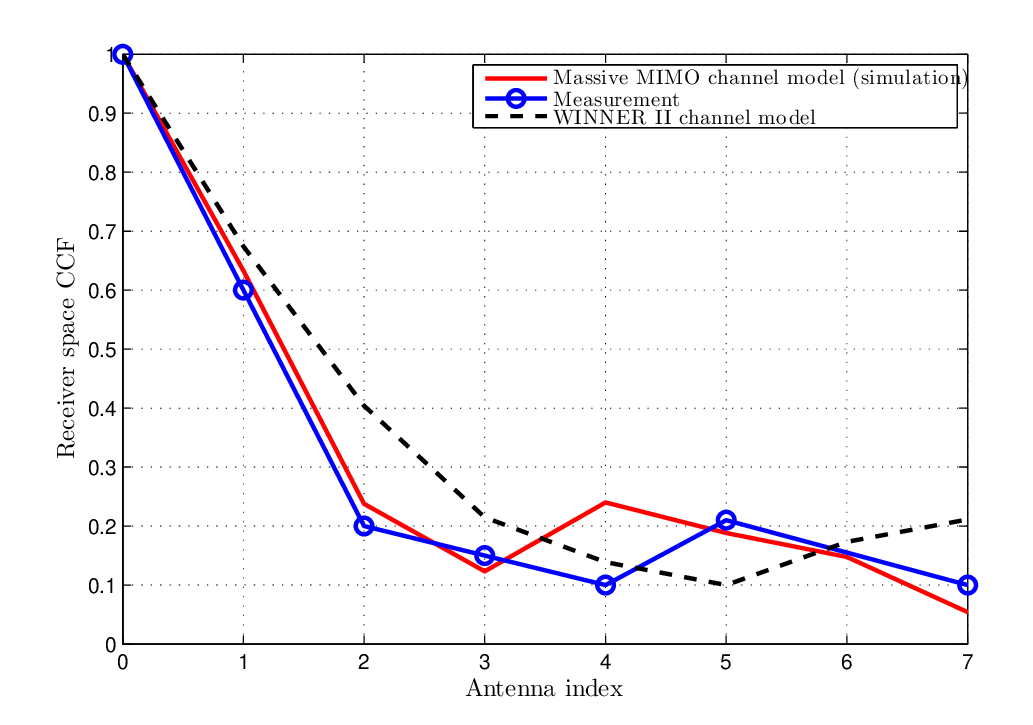}
\centering\caption{The receiver normalized space CCFs of the 3D wideband massive MIMO channel model (simulations), measurement in \cite{2dot6GHz}, and WINNER~II channel model ($f_c=2.6$~GHz \cite{2dot6GHz}, $\|\mathbf{D}\|=200$~m, $M_R=M_T=32$, $\Delta v^R=\Delta v^T=0$~m/s \cite{2dot6GHz}, $|\mathbf{v}^T|=|\mathbf{v}^R|=0$ \cite{2dot6GHz}, $D_c^a=30$~m, $D_c^s=100$~m, $\kappa=-8$~dB, polarized antennas, NLOS).}
\label{fig_MassiveMIMO_SCCF}
\end{figure}

% massive MIMO results
The receiver normalized space CCF of the 3D wideband massive MIMO channel model (simulation) in Table \ref{tab_simulation_parameters}, the measured averaged space CCF in the NLOS scenario in \cite[Fig. 10]{2dot6GHz}, and the space CCF of the WINNER II channel model are compared in Fig. \ref{fig_MassiveMIMO_SCCF}. The horizontal axis of Fig. \ref{fig_MassiveMIMO_SCCF} has been normalized with respect to half wavelength. The measurement in \cite{2dot6GHz} was performed in an outdoor environment with a 128-element polarized virtual linear array covering both LOS and NLOS scenarios. Each polarized antenna pair was also separated by half wavelength. When the antenna index difference is larger than $2$, the correlation coefficients drop to a relatively low level and different sub-channels can be considered as uncorrelated. In this case, the fitting is not important anymore. It is clear that the space CCF of the proposed massive MIMO channel model aligns well with the measured data when the antenna index difference is less than $3$. However, the WINNER II channel model overestimates antenna correlations because cluster evolution was not considered on the array axis.

\begin{figure}
\centering\includegraphics[width=3.3in]{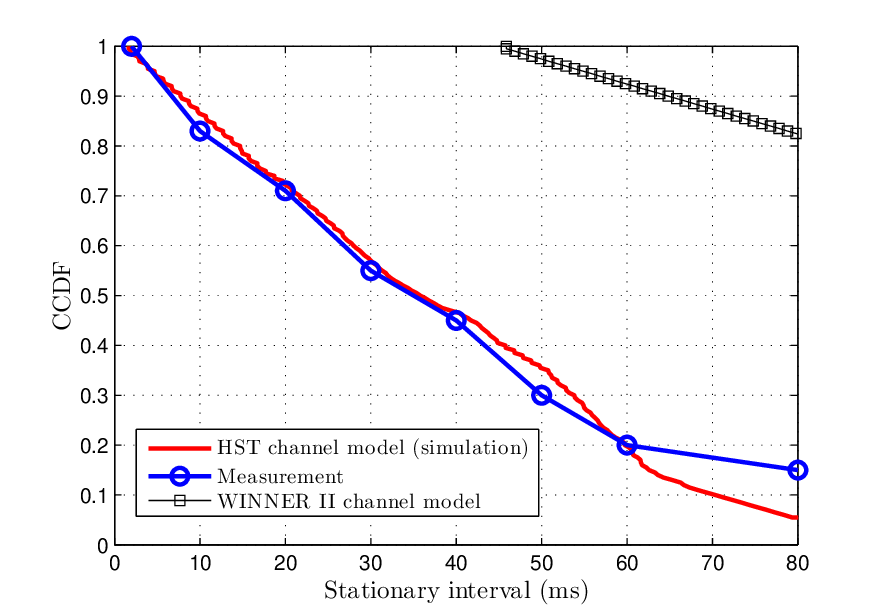}
\centering\caption{The CCDFs of the stationary intervals for the 3D wideband HST conventional MIMO channel model (simulation), measurement in \cite{Chen12}, and WINNER~II channel model ($f_c=932$~MHz \cite{Chen12}, $\|\mathbf{D}\|=200$~m, $M_R=M_T=2$, $\Delta v^R=\Delta v^T=0.5$~m/s, $\|\mathbf{v}^R \|=60$~m/s \cite{Chen12}, $\|\mathbf{v}^T \|=0$~m/s \cite{Chen12}, $D_c^a=50$~m, $D_c^s=100$~m, $\varsigma=7$~s, NLOS).}
\label{fig_HST_Stationary_Interval}
\end{figure}

% M2M and HST
The CCDFs of the stationary intervals of the 3D wideband HST conventional MIMO channel model (with velocity 60~m/s) in Table \ref{tab_simulation_parameters}, the measurement in \cite[Fig. 4(a)]{Chen12} with an omnidirectional antenna, and the WINNER II channel model are shown in Fig. \ref{fig_HST_Stationary_Interval}. The median of the stationary interval is approximately $40$~ms. Clearly, the proposed HST channel model can fit the measurement result very well. The WINNER II channel model has a much higher stationary interval due to the fact that it does not have cluster power evolution in the time domain.

\begin{figure}
\centering\includegraphics[width=3.3in]{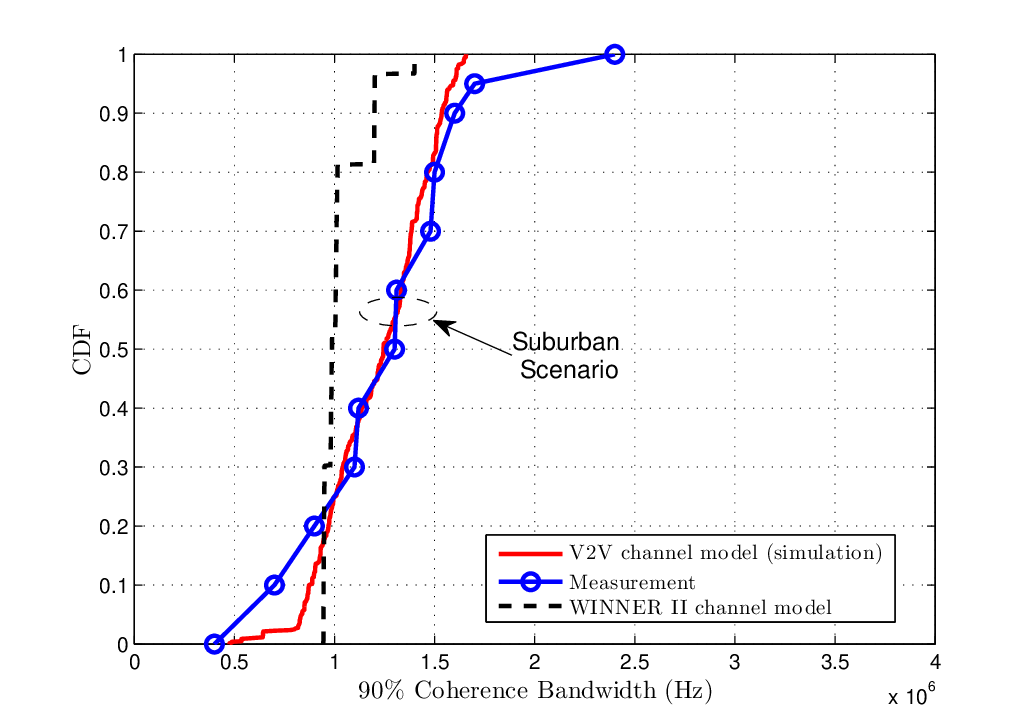}
\centering\caption{The CDFs of $90$\% coherence bandwidths of the 2D wideband V2V conventional MIMO channel model, measurement (suburband scenario) in \cite{Cheng08}, and WINNER~II channel model ($f_c=5.9$GHz \cite{Cheng08}, $\|\mathbf{D}\|=400$m, $\Delta v^R=\Delta v^T=0.5$m/s, $\|\mathbf{v}^T \|=\|\mathbf{v}^R \|=25$m/s, $D_c^a=30$m, $D_c^s=10$m, $\varsigma=5$s, NLOS).}
\label{fig_CDF_90pct_Coherence_Bandwidth_M2M}
\end{figure}

%Fig. \ref{fig_FCF_F2MvsM2M_XPD} illustrates the normalized time-variant FCFs at different time instants. It can be observed that the FCFs are time dependent. Namely, the FCFs of the proposed unified framework are not WSS with respect to time. The coherence bandwidth is approximately $1$ MHz.
The $90$\% coherence bandwidth measures the bandwidth in which the channel can be regarded as flat. The cumulative distribution function (CDF) of $90$\% coherence bandwidth of the 2D wideband V2V conventional MIMO channel model in Table \ref{tab_simulation_parameters} is illustrated in Fig. \ref{fig_CDF_90pct_Coherence_Bandwidth_M2M}, which is also validated by that of the measured suburban V2V channel in \cite[Fig. 5(a)]{Cheng08}. The measurement in \cite{Cheng08} was performed in a suburban scenario with both transmitter and receiver moving in the same direction. The $90$\% coherence bandwidth of the WINNER II channel model has smaller spread because its PDP is not time variant.

\begin{figure}
\centering\includegraphics[width=3.3in]{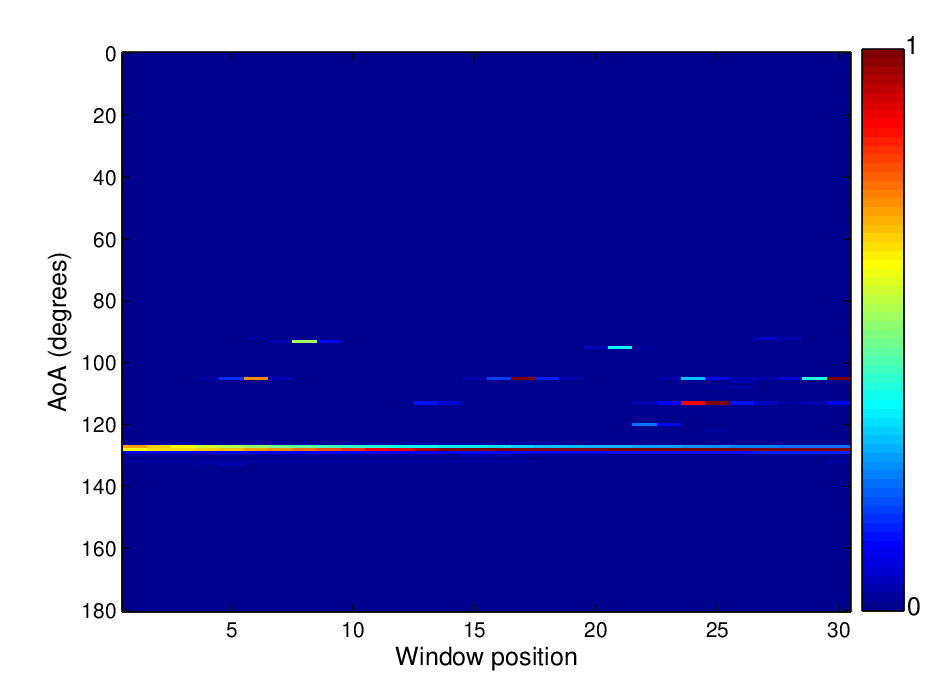}
\centering\caption{A snapshot of the simulated normalized APS at the receiver of the 2D mmWave massive MIMO channel model ($f_c=58$~GHz, $M_T=2$, $M_R=32$, $\|\mathbf{D}\|=6$~m, $\Delta v^R=\Delta v^T=0$~m/s, $|\mathbf{v}^T|=|\mathbf{v}^R|=0$, $D_c^a=30$~m, $D_c^s=100$~m, NLOS).}
\label{fig_APS_mmWave_massiveMIMO}
\end{figure}

A snapshot of the simulated normalized angular power spectrum (APS) at the receiver with a half-wavelength linear array of the 2D mmWave massive MIMO channel model is illustrated in Fig. \ref{fig_APS_mmWave_massiveMIMO}. The APS is estimated with the smooth multiple signal classification (MUSIC) algorithm \cite{WirelessComms}\cite{Shan85} using a sliding window of $3$ consecutive antennas. With these sliding windows, the smooth MUSIC algorithm is able to resolve more AoAs than the conventional MUSIC algorithm. It can be observed that the number of clusters is small because of the high carrier frequency of mmWave. Meanwhile, appearance and disappearance of clusters on the array axis can also be seen due to the massive MIMO antenna array.

\begin{figure}
\centering\includegraphics[width=3.3in]{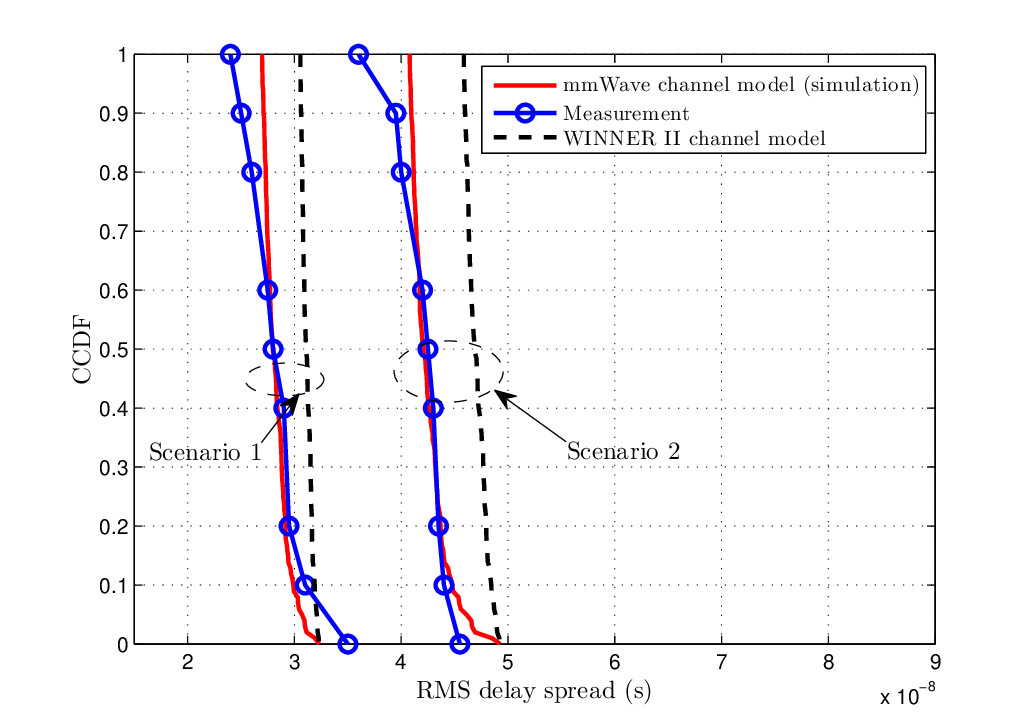}
\centering\caption{The CCDFs of RMS delay spreads of the 3D mmWave conventional MIMO channel model, measurement in \cite{Smulders95}, and WINNER~II channel model ($f_c=58$~GHz \cite{Smulders95}, $\|\mathbf{D}\|=6$~m, $\Delta v^R=\Delta v^T=0$~m/s, $\|\mathbf{v}^T \|=\|\mathbf{v}^R \|=0$~m/s, $D_c^a=30$~m, $D_c^s=100$~m, $\varsigma=7$~s, NLOS).}
\label{fig_mmWave_RMS_Delay_CCDF}
\end{figure}

Fig. \ref{fig_mmWave_RMS_Delay_CCDF} presents the CCDFs of the root mean square (RMS) delay spreads of the 3D mmWave conventional MIMO channel model in Table \ref{tab_simulation_parameters}, the measurements in \cite[Fig. 5]{Smulders95}, and the WINNER~II channel model. The measurements in \cite{Smulders95} were performed in indoor scenarios with omnidirectional antennas in azimuth. Scenario 1 and Scenario 2 are the Room H scenario and Room F scenario in \cite[Fig. 5]{Smulders95}, respectively. It is clear that measurement results for different scenarios in \cite[Fig. 5]{Smulders95} can be fitted properly by the proposed 3D mmWave channel model. The WINNER II channel model does not support mmWave small-scale channel characteristics well as it overestimates the RMS delay spread and has smaller variations. The RMS delay spread of the mmWave channel model is between $20$ ns and $50$ ns.

%%%%%%%%%%%%Numerical Analysis ends%%%%%%%%%%%%%%
%\clearpage
%%%%%%%%%%%Conclusion begins%%%%%%%%%%%%
\section{Conclusions} \label{sec_conclusion_section}
In this paper, a unified framework for 5G wireless small scale fading channel models has been proposed, based on the WINNER II and SV channel models. The proposed general 3D non-stationary 5G channel model can model massive MIMO, V2V, HST, and mmWave communication scenarios, as well as considering time evolution feature of channels and arbitrary antenna array layouts. The modeling of time evolution of channels includes cluster evolution on the time axis, geometrical relationship updates, and evolution of delays and powers of rays. It has been shown that the simulated statistical properties of the proposed general 3D non-stationary 5G GBSM can fit well the corresponding measurements. The proposed general 5G GBSM can be reduced to various simplified channel models by properly setting certain parameters. For future work, applications of the proposed channel model to 5G system simulators and parameter estimations of the proposed 5G channel model from more channel measurements need to be considered. Also, the general 5G GBSM can be further extended by considering cooperative MIMO channel model with multi-link correlations, HST channel model with time-variant Rician factor, HST channel model in tunnel scenarios, time-variant antenna pattern, and the Rician factor evolution in both time and array domains in massive MIMO scenarios.

%%%%%%%%%%%Conclusion ends%%%%%%%%%%%%%%

\appendices
\section{Calculation of Antenna Patterns in (\ref{equ_LOS_component})}\label{Appendix_Antenna_Pattern}
%\begin{figure}
%\centering\includegraphics[width=2.5in]{LCS_GCS.eps}
%\centering\caption{Diagram of LCS and GCS in the 3D space.}
%\label{fig_LCS_GCS}
%\end{figure}
In a 3D space, the geometry of antenna pattern depends on the orientation of antennas. The LCS is obtained by a sequence of rotations from the GCS \cite{36873}. First, a rotation of $\alpha^T$ about $x_\mathrm{G}$ axis is operated. Second, a rotation of $\beta^T$ about the new $y_\mathrm{G}$ axis is operated. Third, a rotation of $\gamma^T$ about the new $z_\mathrm{G}$ axis is operated. These three operations can be expressed as (\ref{equ_antenna_rotation}) \cite{36873}.
\begin{figure*}[!t]
% ensure that we have normalsize text
\normalsize
% Store the current equation number.
% Set the equation number to one less than the one
% desired for the first equation here.
% The value here will have to changed if equations
% are added or removed prior to the place these
% equations are referenced in the main text.
\begin{align}
\mathbf{R}=\begin{bmatrix}
\cos\gamma^{T} & -\sin\gamma^{T}  &0 \\
\sin\gamma^{T} & \cos\gamma^{T}  &0 \\
0 &0  &1
\end{bmatrix}
\begin{bmatrix}
\cos\beta^{T} & 0 &\sin\beta^{T} \\
 0&1  &0 \\
-\sin\beta^{T} & 0 & \cos\beta^{T}
\end{bmatrix}
\begin{bmatrix}
1 &0  &0 \\
0 & \cos\alpha^{T}  & -\sin\alpha^{T}  \\
0 &\sin\alpha^{T}  & \cos\alpha^{T}
\end{bmatrix}.
\label{equ_antenna_rotation}
\end{align}
% Restore the current equation number.
% IEEE uses as a separator
\hrulefill
% The spacer can be tweaked to stop underfull vboxes.
\vspace*{4pt}
\end{figure*}

Let $\mathbf{a}$ and $\mathbf{b}$ be positions vectors in the GCS, and let $\left(\tilde{x},\tilde{y},\tilde{z} \right)^{\mathrm{T}}$ be the coordinates of $\mathbf{a}-\mathbf{b}$ in the LCS. Then, $\left(\tilde{x},\tilde{y},\tilde{z} \right)^{\mathrm{T}}=\mathbf{R}(\mathbf{a}-\mathbf{b})$. Let $w(x,y)$ be the four-quadrant inverse tangent function \cite{Glisson11} of $x$ and $y$, and let $\tilde{\theta}=w(\tilde{y},\tilde{x})$ and $\tilde{\phi}=w(\tilde{z},\sqrt{\tilde{x}^2+\tilde{y}^2})$. Then, the antenna patterns in (9) can be computed as $F_H(\mathbf{a},\mathbf{b})=G(\tilde{\theta},\tilde{\phi})\cos\tilde{\theta}$ and $F_V(\mathbf{a},\mathbf{b})=G(\tilde{\theta},\tilde{\phi})\sin\tilde{\theta}$.
%\begin{align}
%F_H(\mathbf{a},\mathbf{b})=G(\tilde{\theta},\tilde{\phi})\cos\tilde{\theta}\\
%F_V(\mathbf{a},\mathbf{b})=G(\tilde{\theta},\tilde{\phi})\sin\tilde{\theta}.
%\end{align}
% $F_H(\mathbf{a},\mathbf{b})=G(\tilde{\theta},\tilde{\phi})\cos\tilde{\theta}$ and $F_V(\mathbf{a},\mathbf{b})=G(\tilde{\theta},\tilde{\phi})\sin\tilde{\theta}$.
In this paper, dipole antennas are assumed at the transmitter side. In this case \cite{Elliott03},  $G(\tilde{\theta},\tilde{\phi})=\sqrt{1.64}\frac{\cos\left(\frac{\pi}{2}\cos\tilde{\phi} \right)}{\sin\tilde{\phi}}.$
%\begin{align}
%G(\tilde{\theta},\tilde{\phi})=\sqrt{1.64}\frac{\cos\left(\frac{\pi}{2}\cos\tilde{\phi} \right)}{\sin\tilde{\phi}}.
%\end{align}
%$G(\tilde{\theta},\tilde{\phi})=\sqrt{1.64}\frac{\cos\left(\frac{\pi}{2}\cos\tilde{\phi} \right)}{\sin\tilde{\phi}}$.
At the receiver side, calculation follows a similar procedure. Omnidirectional antennas are assumed at the receiver side. In this case, $G(\tilde{\theta},\tilde{\phi})=1$. Both antenna patterns can be replaced by the actual antennas used. The rotation angles $\alpha^T$, $\alpha^R$, $\beta^T$, $\beta^R$, $\gamma^T$, and $\gamma^R$ are all set as $\frac{\pi}{15}$ for simplicity, which can be modified according to realistic settings.

\section{Calculation of (\ref{equ_Pn_evolution})}\label{Appendix_Cluster_Power}
Let us consider the time interval between $t$ and $t+\Delta t$, where $\Delta t$ is small. Then, the ray mean power difference $\Delta \tilde{ P}_{n,m_n}(t)$ between these two time instants is computed as
\begin{align}
\Delta \tilde{ P}_{n,m_n}(t)=\tilde{ P}_{n,m_n}(t+\Delta t)-\tilde{ P}_{n,m_n}(t).
\label{equ_Delta_P}
\end{align}
With the assumption of the inverse power law, the ray mean power is inversely proportional to the $\eta$th ($\eta>1$) power of travel distance, i.e.,
$\tilde{ P}_{n,m_n}(t)= \frac{\mathcal{C}}{\left[\tau_n(t)+\tau_{n,m_n} \right]^{\eta}c^\eta}$,
%\begin{align}
%\tilde{ P}_{n,m_n}(t)= \frac{\mathcal{C}}{\left[\tau_n(t)+\tau_{n,m_n} \right]^{\eta}c^\eta}
%\end{align}
where $\mathcal{C}$ is a constant. The derivative of $\tilde{ P}_{n,m_n}(t)$ with respect to $t$ is obtained by
\begin{align}
\frac{\Delta \tilde{ P}_{n,m_n}(t)}{\Delta t}&= \left(\frac{\mathcal{C}}{\left[\tau_n(t)+\tau_{n,m_n} \right]^{\eta}c^\eta} \right)'\nonumber\\
&=-\eta\frac{\mathcal{C}}{\left[\tau_n(t)+\tau_{n,m_n} \right]^{\eta+1}c^{\eta}}\left(\tau_n(t) \right)'\nonumber\\
&=-\eta\frac{\mathcal{C}}{\left[\tau_n(t)+\tau_{n,m_n} \right]^{\eta+1}c^{\eta}}\frac{\left[\tau_n(t+\Delta t)-\tau_n(t)\right]}{\Delta t}.
\label{equ_Delta_P_over_Delta_t}
\end{align}
Thus, the ray mean power evolution in terms of time can be derived as
\begin{align}
\frac{\Delta \tilde{ P}_{n,m_n}(t)}{\tilde{ P}_{n,m_n}(t)}=\frac{-\eta }{\tau_n(t)+\tau_{n,m_n} } \frac{\left[\tau_n(t+\Delta t)-\tau_n(t)\right]}{\Delta t}\Delta t.
\end{align}
It follows that
\begin{align}
\tilde{ P}_{n,m_n}(t+\Delta t)=\tilde{ P}_{n,m_n}(t)\frac{(\eta+1)\tau_n(t)-\eta\tau_n(t+\Delta t)+\tau_{n,m_n}}{\tau_n(t)+\tau_{n,m_n}}.
\end{align}
For simplicity, in this paper we use $\eta=2$ following the inverse square law. Then, (\ref{equ_Pn_evolution}) is obtained.

\section{Parameter Estimation Procedure of the Proposed Model}\label{Appendix_ParaEstimation}

A typical parameter estimation method can be found in \cite{Roivainen17}. This paper follows a similar procedure as in \cite{Roivainen17},  i.e., estimating parameters minimizing the difference in statistical properties between the model and measurement. The parameter estimation procedure in this paper directly estimates parameters to fit the statistic properties of the channel. The parameters of the proposed model can be divided into three categories, i.e., parameters determined via reasonable assumptions, parameters determined via parameter estimation methods, and parameters randomly generated. The categories of parameters are listed in Table \ref{tab_para_estimation_appendix}. Let $\mathsf{P}$ be the set of parameters to be estimated, i.e., $\mathsf{P}=\left\lbrace  \mathrm{std}\left[\phi_{n}^{A} \right],\mathrm{std}\left[\phi_{n}^{E}\right], \mathrm{std}\left[\varphi_{n}^{A}\right], \mathrm{std}\left[\varphi_{n}^{E}\right], D^a_c, D^s_c, \varsigma \right\rbrace$
%\begin{align}
%\mathsf{P}=\left\lbrace \bar{\phi}_{n}^{A}, \bar{\phi}_{n}^{E}, \bar{\varphi}_{n}^{A}, \bar{\varphi}_{n}^{E},  \mathrm{std}\left[\phi_{n}^{A} \right],\mathrm{std}\left[\phi_{n}^{E}\right], \mathrm{std}\left[\varphi_{n}^{A}\right], \mathrm{std}\left[\varphi_{n}^{E}\right], D^a_c, D^s_c, \varsigma \right\rbrace
%\label{equ_para_set0}
%\end{align}
%\begin{align}
%\mathsf{P}=\left\lbrace  \mathrm{std}\left[\phi_{n}^{A} \right],\mathrm{std}\left[\phi_{n}^{E}\right], \mathrm{std}\left[\varphi_{n}^{A}\right], \mathrm{std}\left[\varphi_{n}^{E}\right], D^a_c, D^s_c, \varsigma \right\rbrace
%\label{equ_para_set0}
%\end{align}
and ${\mathsf{f}}(\mathsf{P})$ be a statistical property with channel impulse responses generated using $\mathsf{P}$. The estimated parameters $\hat{\mathsf{P}}$ can be obtained via optimization methods, e.g., exhaustive search, by minimizing $\hat{\mathsf{P}}=\arg\min_{\mathsf{P}}|\hat{{\mathsf{f}}}-\mathsf{f}(\mathsf{P})|^2$ where $\hat{\mathsf{f}}$ is the measured statistical property. This can be summarized into a number of steps.
\begin{enumerate}[Step 1:]
\item Define target threshold $\epsilon_\mathrm{T}$ and initialize parameters $\hat{\mathsf{P}}$.
\item Generate channel coefficients with parameters $\hat{\mathsf{P}}$.
\item Calculate the statistical property $\mathsf{f}(\hat{\mathsf{P}})$ using the generated channel coefficients.
\item Compute the error between the model and measurement $\epsilon=|\hat{\mathsf{f}}-\mathsf{f}(\hat{\mathsf{P}})|^2$.
\item If $\epsilon\leqslant \epsilon_\mathrm{T}$, output $\hat{\mathsf{P}}$; Otherwise, generate a new set of $\hat{\mathsf{P}}$ and go to Step 2.

\end{enumerate}
\begin{table*}[!htbp]
\caption{Details of parameter estimation.}
\center
\footnotesize
\begin{tabular}{|c|c|c|}
\hline
%\multicolumn{3}{|c|}{wideband 3D massive MIMO} & \multicolumn{3}{c|}{HST wideband 3D conventional MIMO} \\ \hline
   Parameters    &  Mean     & Standard deviation      \\ \hline
   $M_n$    &   Determined via assumption    &    0     \\ \hline
   $\tau_n$    &    Randomly generated via WINNER II   &    Randomly generated via WINNER II         \\ \hline
   $\tau_{m_n}$, $D_n^T$, $D_n^R$    &   Determined via assumption     &    Determined via assumption         \\ \hline

%       &   Determined via assumption    &   Determined via assumption         \\ \hline
%      &    Determined via assumption   &   Determined via assumption     \\ \hline
    $\phi_{n}^{A}$, $\phi_{n}^{E}$, $\varphi_{n}^{A}$, $\varphi_{n}^{E}$   &   Randomly generated via WINNER II     &   Determined via parameter estimation    \\ \hline
%       &    Randomly generated via WINNER II     &   Determined via parameter estimation         \\ \hline
%       &  Randomly generated via WINNER II    &   Determined via parameter estimation     \\ \hline
%       &   Randomly generated via WINNER II    &  Determined via parameter estimation     \\ \hline

   $D^a_c$   &    Determined via parameter estimation   &    0 m       \\ \hline
   $D^s_c$    &   Determined via parameter estimation    &    0 m     \\ \hline
   $\varsigma$    &   Determined via parameter estimations    &   0 s            \\ \hline
%$D^R_n$   &    Determined via assumption   &    Determined via assumption       \\ \hline
%   $D^T_n$    &   Determined via assumption    &   Determined via assumption     \\ \hline
\end{tabular}
\label{tab_para_estimation_appendix}
\end{table*}
\ifCLASSOPTIONcaptionsoff
  \newpage
\fi



%\bibliographystyle{IEEEtranTCOM}
%%%%%%%%%%%Bibliography begins%%%%%%%%%%%%

%%%%%%%%%%%Bibliography ends%%%%%%%%%%%%%%

\begin{IEEEbiography}[{\includegraphics[width=1.1in,height=1.3in,clip,keepaspectratio] {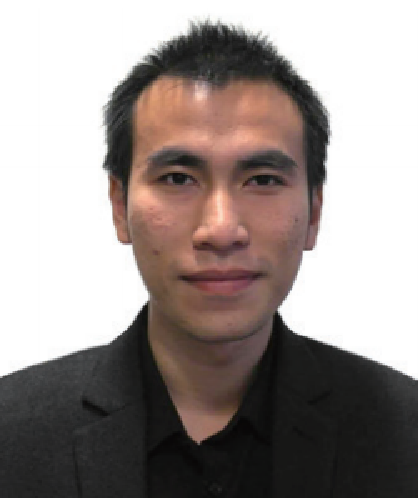}}]
{Shangbin Wu} received his B.S. degree incommunication engineering from South China Normal University, Guangzhou, China, in 2009, M.Sc. degree in wireless communications with distinction from University of Southampton, Southampton, UK, in 2010, and Ph.D. degree in electrical engineering from Heriot-Watt University, Edinburgh, UK in 2015. 

From 2010 to 2011, he worked as a LTE R\&D engineer responsible for LTE standardization and system
level simulation in New Postcom Equipment Ltd., Guangzhou, China. From October 2011 to August 2012, he was with Nokia Siemens Network, where he worked as a LTE algorithm specialist, mainly focusing on LTE radio resource management algorithm design and system level simulations. He has been with Samsung R\&D Institute UK as a 5G researcher since November 2015.
\end{IEEEbiography}

\begin{IEEEbiography}[{\includegraphics[width=1.1in,height=1.3in,clip,keepaspectratio] {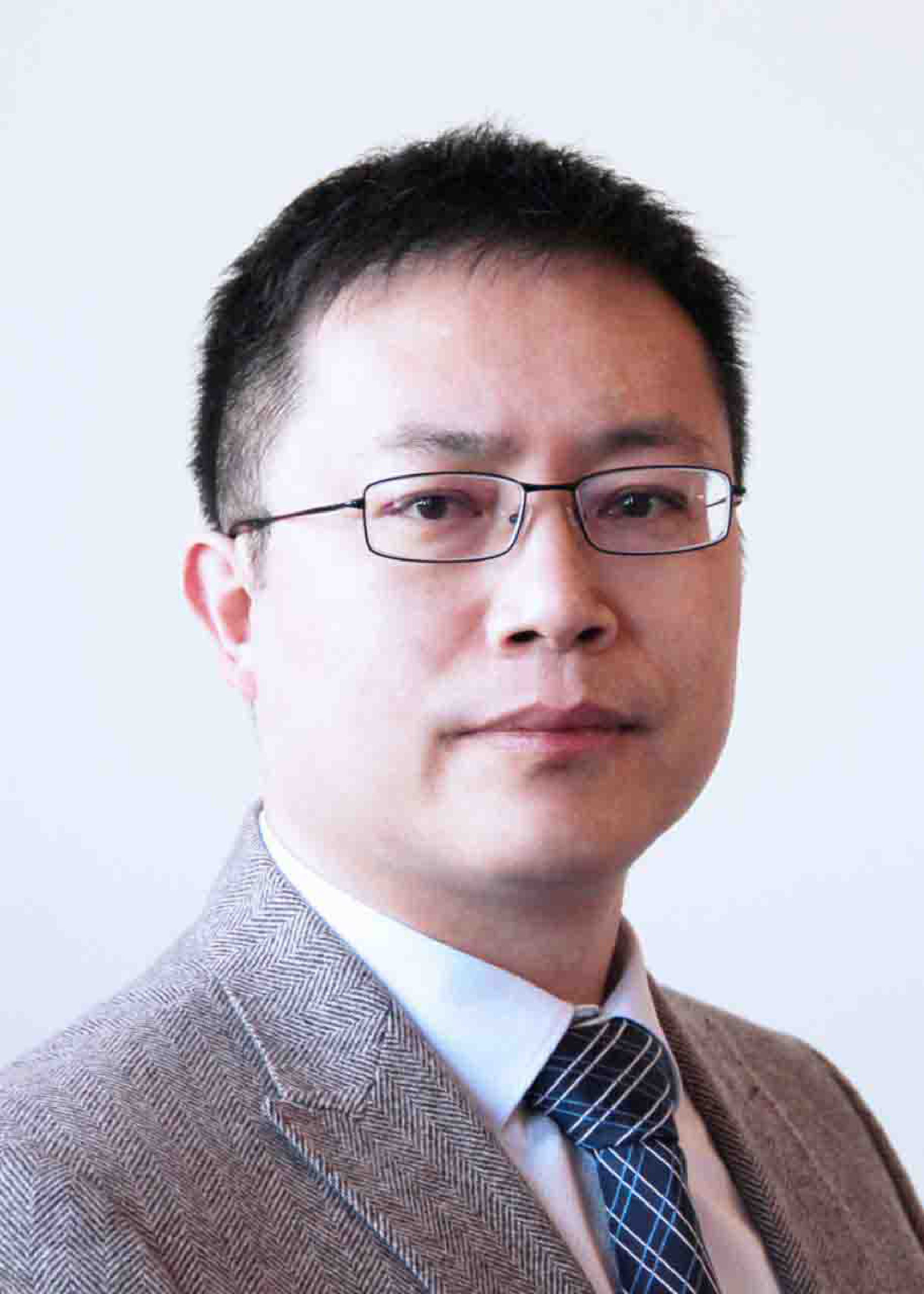}}]
{Cheng-Xiang Wang (S'01-M'05-SM'08-F'17)} received the BSc and MEng degrees in Communication and Information Systems from Shandong University, China, in 1997 and 2000, respectively, and the PhD degree in Wireless Communications from Aalborg University, Denmark, in 2004.

He was a Research Fellow with the University of Agder, Grimstad, Norway, from 2001 to 2005, a Visiting Researcher with Siemens AG Mobile Phones, Munich, Germany, in 2004, and a Research Assistant with the Hamburg University of Technology, Hamburg, Germany, from 2000 to 2001. He has been with Heriot-Watt University, Edinburgh, U.K., since 2005, where he was promoted to a Professor in 2011. He is also an Honorary Fellow of the University of Edinburgh, U.K., and a Chair/Guest Professor of Shandong University and Southeast University, China. He has authored 2 books, one book chapter, and over 300 papers in refereed journals and conference proceedings. His current research interests include wireless channel modeling and (B)5G wireless communication networks, including green communications, cognitive radio networks, high mobility communication networks, massive MIMO, millimetre wave communications, and visible light communications.

Prof. Wang is a Fellow of the IET and HEA, and a member of the EPSRC Peer Review College. He served or is currently serving as an Editor for nine international journals, including the IEEE TRANSACTIONS ON VEHICULAR TECHNOLOGY since 2011, the IEEE TRANSACTIONS ON COMMUNICATIONS since 2015, and the IEEE TRANSACTIONS ON WIRELESS COMMUNICATIONS from 2007 to 2009. He was the leading Guest Editor of the IEEE JOURNAL ON SELECTED AREAS IN COMMUNICATIONS, Special Issue on Vehicular Communications and Networks. He is also a Guest Editor of the IEEE JOURNAL ON SELECTED AREAS IN COMMUNICATIONS, Special Issue on Spectrum and Energy Efficient Design of Wireless Communication Networks and Special Issue on Airborne
Communication Networks, and the IEEE TRANSACTIONS ON BIG DATA, Special Issue on Wireless Big Data. He served or is serving as a TPC Member, TPC Chair, and General Chair of over 80 international conferences. He received nine Best Paper Awards from the IEEE Globecom 2010, the IEEE
ICCT 2011, ITST 2012, the IEEE VTC 2013, IWCMC 2015, IWCMC 2016, the IEEE/CIC ICCC 2016, and the WPMC 2016. He is recognized as Web of Science 2017 Highly Cited Researcher.
\end{IEEEbiography}

\begin{IEEEbiography}[{\includegraphics[width=1.1in,height=1.3in,clip,keepaspectratio] {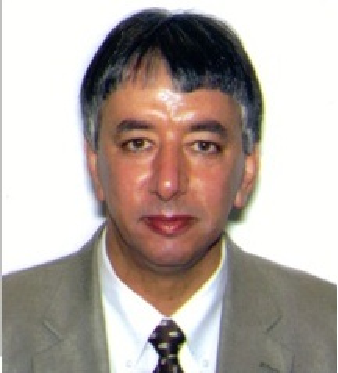}}]
{el-Hadi M. Aggoune (M'83-SM'93)} received the M.S. and Ph.D. degrees in electrical engineering from the University of Washington (UW), Seattle,
WA, USA. He taught graduate and undergraduate courses in electrical engineering at many universities in the United States and abroad. He served at many academic ranks including Endowed Chair Professor. He is listed as Inventor in two patents assigned to the Boeing Company, USA and the Sensor Networks and Cellular Systems (SNCS) Research Center, University of Tabuk, Saudi Arabia. His research is referred to in many patents, including patents assigned to ABB, Switzerland and EPRI, USA. He authored many papers in IEEE and other journals and conferences. Dr. Aggoune is serving on many technical committees for conferences worldwide as well as reviewer for many journals. One of his Labs won the Boeing Supplier Excellence Award. Dr. Aggoune won the IEEE Professor of the Year Award, UW Branch. He is a Professional Engineer registered in the State of Washington. He is currently serving as Professor and the Director of SNCS Research Center, University of Tabuk. His research interests include wireless sensor networks, energy systems, and scientific visualization.
\end{IEEEbiography}

\begin{IEEEbiography}[{\includegraphics[width=1.1in,height=1.3in,clip,keepaspectratio] {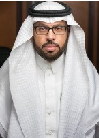}}]
{Mohammed M. Alwakeel (SM'14)} was born in Tabuk, Saudi Arabia. He received his B.S. and M. S Degrees from King Saud University, Riyadh, Saudi Arabia, and his Ph.D. Degree in Electrical Engineering from Florida Atlantic University, Boca Raton, Florida. He served as Communications Network Manager at the Saudi National Information Center in Riyadh. He served as faculty member at King Abdulaziz University and then Associate Professor and Dean of the Computers and Information Technology College at the University of Tabuk, Tabuk, Saudi Arabia. After that, he was a full professor at Computers and Information Technology College, and the Vice Rector for Development and Quality at the University of Tabuk. Currently, he is a member of Alshura Council (The Consultative Council of the Kingdom of Saudi Arabia). His current research interests include teletraffic analysis, mobile satellite communications, and sensor networks and cellular systems.
\end{IEEEbiography}

\begin{IEEEbiography}[{\includegraphics[width=1.1in,height=1.3in,clip,keepaspectratio] {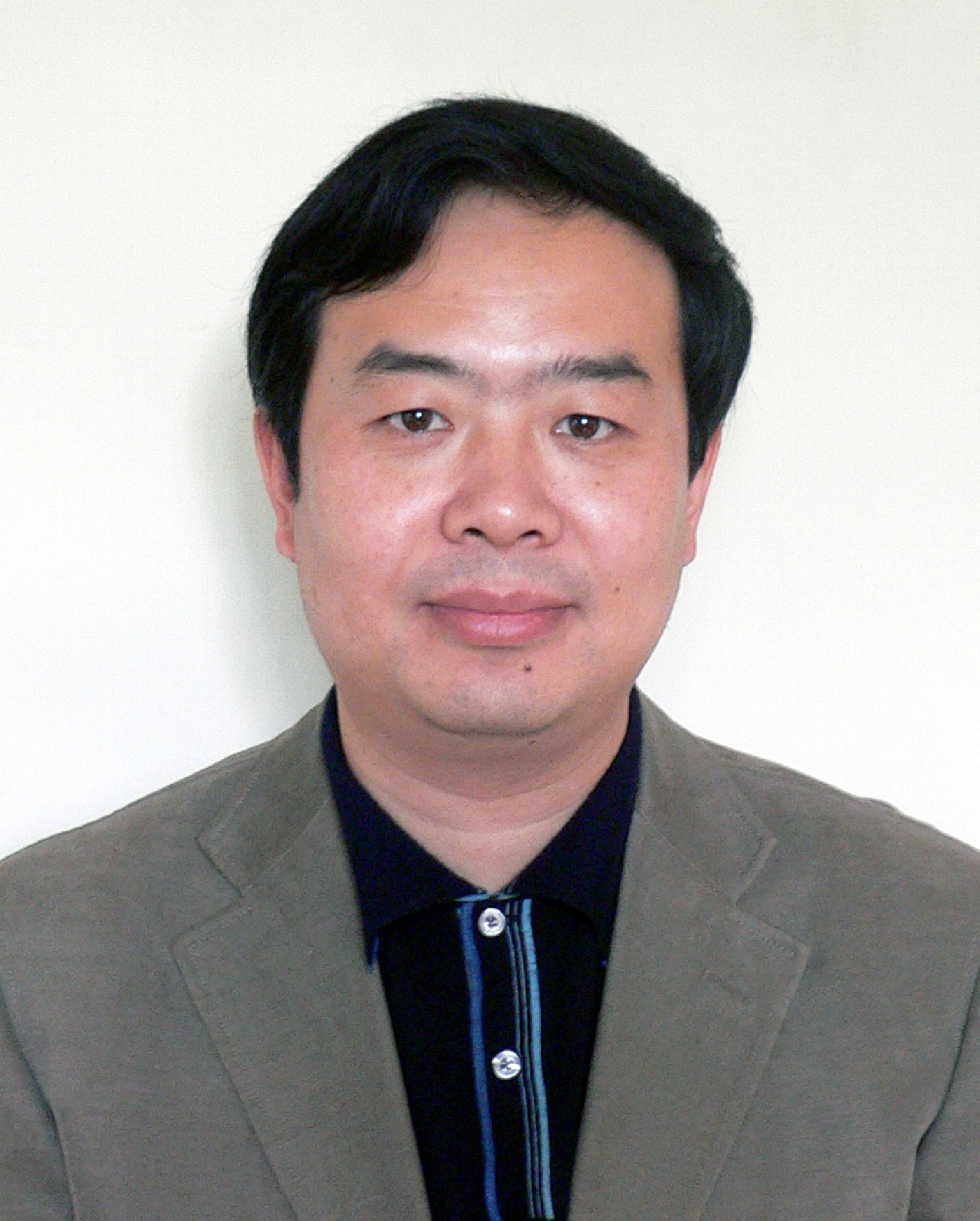}}]
{Xiaohu You (F'12)} received his Ph.D. Degrees from Southeast University, Nanjing, China, in Electrical Engineering in 1988. Since 1990, he has been working with National Mobile Communications Research Laboratory at Southeast University, where he is currently a professor and the director. He has contributed over 150 IEEE journal papers in the areas of signal processing and wireless communications. From 1998 to 2016, he was the Principal Expert of the 3G, 4G and 5G Research Projects under China National 863 High-Tech Program, respectively. 
Professor You served as the general chairs of IEEE WCNC 2013 and IEEE VTC 2016. He was the recipient of the China National 1st Class Invention Prize in 2011, and was selected as IEEE Fellow in same year.

\end{IEEEbiography}

\end{document}